\documentclass[twocolumn,superscriptaddress,prl,amsmath,amssymb,floatfix,longbibliography,nofootinbib]{revtex4-2}
\usepackage[latin9]{inputenc}
\usepackage{color}
\usepackage{mathtools}
\usepackage{amsmath}
\usepackage{graphicx}
\usepackage{esint}
\usepackage[unicode=true,
 bookmarks=false,
 breaklinks=false,pdfborder={0 0 1},backref=false,colorlinks=true]
 {hyperref}

\makeatletter

\usepackage{dcolumn}
\usepackage{bm}
\usepackage[usenames,dvipsnames]{xcolor}
\usepackage{bbm}
\usepackage{braket}
\usepackage{comment}
\usepackage{upgreek}
\usepackage{soul}
\usepackage[normalem]{ulem}
\usepackage{floatrow}
\usepackage{orcidlink}

\newcommand{\EqualContribution}{These authors contributed equally to this work}

\newcommand{\beginsupplement}{
    \setcounter{section}{0}
    \renewcommand{\thesection}{S\arabic{section}}
    \setcounter{equation}{0}
    \renewcommand{\theequation}{S\arabic{equation}}
    \setcounter{table}{0}
    \renewcommand{\thetable}{S\arabic{table}}
    \setcounter{figure}{0}
    \renewcommand{\thefigure}{S\arabic{figure}}
    \newcounter{SMfig}
    \renewcommand{\theSMfig}{S\arabic{SMfig}}}

\makeatother

\begin{document}
\global\long\def\sgn{\mathrm{sgn}}%
\global\long\def\ket#1{\left|#1\right\rangle }%
\global\long\def\bra#1{\left\langle #1\right|}%
\global\long\def\sp#1#2{\langle#1|#2\rangle}%
\global\long\def\abs#1{\left|#1\right|}%
\global\long\def\avg#1{\langle#1\rangle}%

\title{Scaling tunnelling noise in the fractional quantum Hall effect tells
about renormalization and breakdown of chiral Luttinger liquid}
\author{Noam Schiller\,\orcidlink{0000-0001-7809-5327}}
\thanks{\EqualContribution}
\affiliation{{\small{}Department of Condensed Matter Physics, Weizmann Institute
of Science, Rehovot 7610001, Israel}}
\author{Tomer Alkalay\,\orcidlink{0009-0003-2380-9179}}
\thanks{\EqualContribution}
\affiliation{{\small{}Braun Center for Submicron Research, Department of Condensed
Matter Physics, Weizmann Institute of Science, Rehovot 7610001, Israel}}
\author{Changki Hong\,\orcidlink{0000-0002-3801-1659}}
\thanks{\EqualContribution}
\affiliation{{\small{}Braun Center for Submicron Research, Department of Condensed
Matter Physics, Weizmann Institute of Science, Rehovot 7610001, Israel}}
\author{Vladimir Umansky\,\orcidlink{0000-0001-5727-2064}}
\affiliation{{\small{}Braun Center for Submicron Research, Department of Condensed
Matter Physics, Weizmann Institute of Science, Rehovot 7610001, Israel}}
\author{Moty Heiblum\,\orcidlink{0000-0002-9331-5022}}
\affiliation{{\small{}Braun Center for Submicron Research, Department of Condensed
Matter Physics, Weizmann Institute of Science, Rehovot 7610001, Israel}}
\author{Yuval Oreg\,\orcidlink{0000-0001-8753-8468}}
\affiliation{{\small{}Department of Condensed Matter Physics, Weizmann Institute
of Science, Rehovot 7610001, Israel}}
\author{Kyrylo Snizhko\,\orcidlink{0000-0002-7236-6779}}
\affiliation{{\small{}Department of Condensed Matter Physics, Weizmann Institute
of Science, Rehovot 7610001, Israel}}
\affiliation{{\small{}Institute for Quantum Materials and Technologies, Karlsruhe
Institute of Technology, 76021 Karlsruhe, Germany}}
\affiliation{Univ. Grenoble Alpes, CEA, Grenoble INP, IRIG, PHELIQS, 38000 Grenoble,
France}
\date{\today}
\begin{abstract}
The fractional quantum Hall (FQH) effect provides a paradigmatic example
of a topological phase of matter. FQH edges are theoretically described
via models belonging to the class of chiral Luttinger liquid (CLL)
theories \citep{wen_quantum_2007}. These theories predict exotic
properties of the excitations, such as fractional charge and fractional
statistics. Despite theoretical confidence in this description and
qualitative experimental confirmations, quantitative experimental
evidence for CLL behaviour is scarce. In this work, we study tunnelling
between edge modes in the quantum Hall regime at the filling factor
$\nu=1/3$. We present measurements at different system temperatures
and perform a novel scaling analysis of the experimental data, originally
proposed in Ref.~\citep{schiller_extracting_2022}. Our analysis
shows clear evidence of CLL breakdown --- above a certain energy
scale. In the low-energy regime, where the scaling behaviour holds,
we extract the property called the \emph{scaling dimension} and find
it heavily renormalized compared to naïve CLL theory predictions.
Our results show that decades-old experiments contain a lot of previously
overlooked information that can be used to investigate the physics
of quantum Hall edges. In particular, we open a road to \textit{quantitative
experimental} studies of (a) scaling dimension renormalization in
quantum point contacts and (b) CLL breakdown mechanisms at an intermediate
energy scale, much smaller than the bulk gap. 
\end{abstract}
\maketitle

\section{Introduction}

Transport measurements of edge modes have long been used to infer
topological properties of non-trivial phases of matter. Such inference
requires an effective theoretical description of the edge modes, related
to the underlying bulk phase through the principle of bulk-boundary
correspondence. This approach has been particularly fruitful in the
case of the quantum Hall (QH) effect --- the earliest known and one
of the most-studied topological phases of matter. The widely accepted
low energy theoretical description of integer and fractional QH edges
is that of a chiral Luttinger liquid (CLL) \citep{wen_quantum_2007}.
It is important to emphasize that CLL is not a model, but a \emph{framework}
that encompasses numerous specific models of QH edges, see \citep[Appendix A in][]{schiller_extracting_2022}. 

Beyond extraction of topological invariants, such as the electric
and thermal conductance, transport experiments with QH edges have
been successful in finding properties of the elementary excitations
supported by the system \citep{stern_anyons_2008}, colloquially known
as quasiparticles. In the case of fractional QH effect, quasiparticles
have been predicted to carry fractional charge \citep{laughlin_anomalous_1983}
and obey anyonic \citep{arovas_fractional_1984}, and in some cases
even non-Abelian \citep{moore_nonabelions_1991}, exchange statistics.
Transport experiments have confirmed the fractional charge of quasiparticles
in several FQH states \citep{de-picciotto_direct_1997,saminadayar_observation_1997,dolev_observation_2008}.
More recently, evidence of the fractional statistics has been observed
\citep{nakamura_direct_2020,bartolomei_fractional_2020,lee_partitioning_2022,Ruelle2023,Glidic2023}.

A crucial tool in obtaining these single-quasiparticle properties
has been partitioning the edge using a quantum point contact (QPC).
Bringing two edges close together, a QPC allows tunnelling of quasiparticles
from one edge to another. The observable outcomes of such tunnelling,
such as the current and its fluctuations (``noise''), are intimately
related to the quantum nature and discreteness of the quasiparticles.
Such partitioning of a Luttinger liquid is a well researched problem
\citep{Wen1991,kane_transport_1992,kane_transmission_1992,Chamon1993,kane_nonequilibrium_1994},
and an exact solution for the conductance through a QPC connecting
two identical Laughlin FQH states (filling factor $\nu=1/m$ for odd
$m$) was presented some thirty years ago \citep{fendley_exact_1994,fendley_exact_1995}.

However, puzzles remain, in the form of disagreements between the
exact solution and experimental observations. Most strikingly, the
hallmark of partitioned Luttinger liquids --- a power law current--voltage
dependence, $I\propto V^{4\delta-1}$ \citep{kane_transport_1992,kane_transmission_1992}
--- is notably absent from experiments \citep{Heiblum2006}. The
power-law exponent is defined by the scaling dimension, $\delta$,
which in turn is related to the zero-temperature time correlations
of the quasiparticles, $\langle\psi^{\dagger}(\tau)\psi(0)\rangle\propto\tau^{-2\delta}$.\footnote{Some literature defines the scaling dimension as $\langle\psi^{\dagger}(\tau)\psi(0)\rangle\propto\tau^{-\delta}$,
with the corresponding prediction $I\propto V^{2\delta-1}$. In this
case, the scaling dimension is often denoted as $g$.} Several causes, including electrostatic interactions \citep{Pryadko2000,papa_interactions_2004,yang_influence_2013},
$1/f$-noise \citep{braggio_environmental_2012} or neutral modes
\citep{rosenow_nonuniversal_2002,ferraro_relevance_2008}, can lead
to deviation of the scaling dimension from pristine theoretical values,
perhaps contributing to this long-standing discrepancy. Some experiments
have found good agreement with Luttinger liquid theory by subtracting
a background conductance \citep{Roddaro2003,Roddaro2004,radu_quasi-particle_2008,Lin2012,rossler_experimental_2014}
--- a heuristic method that is not backed by theory.

\begin{figure}
\centering{}\includegraphics[width=1\columnwidth]{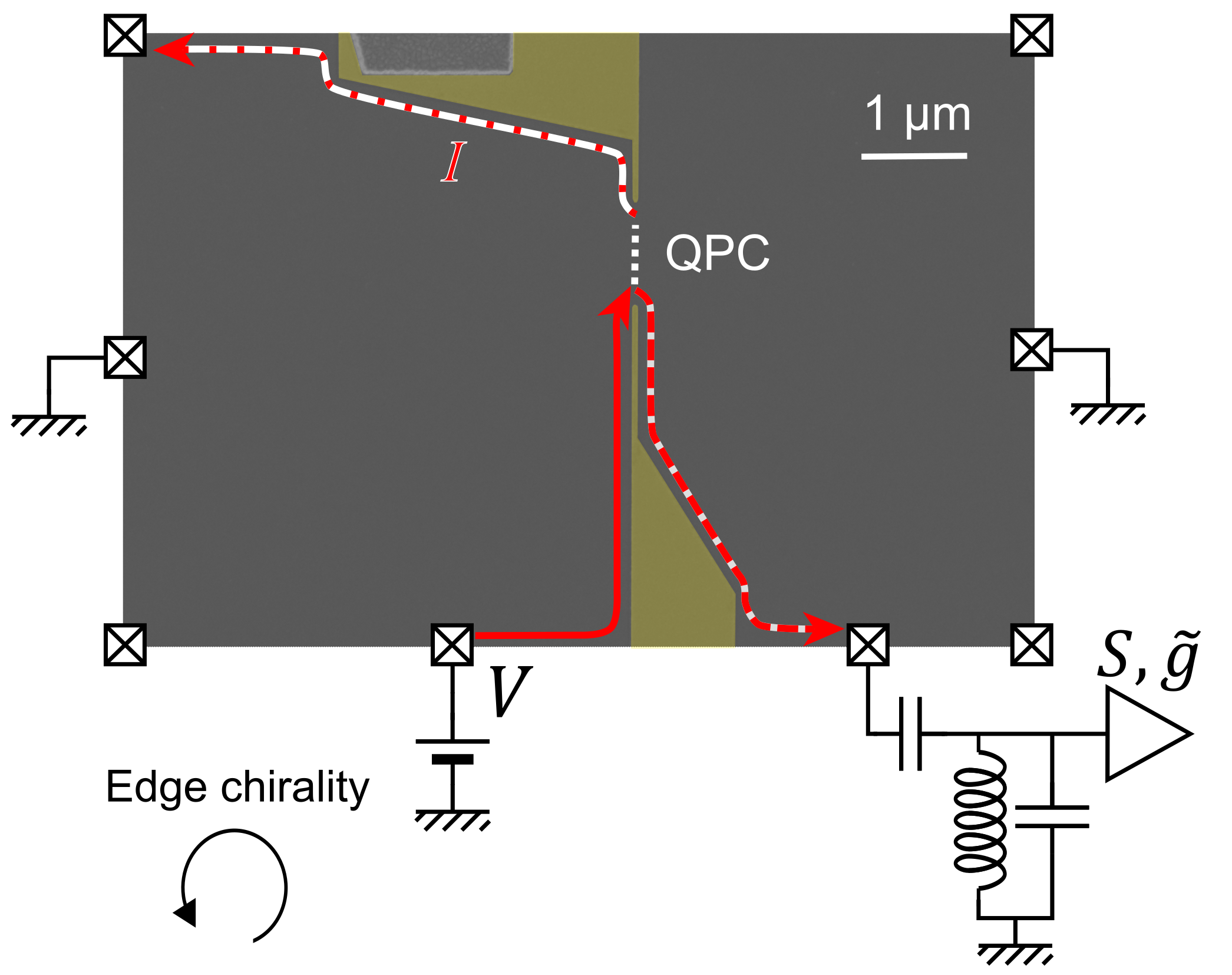}\caption{The experimental setup. False-coloured scanning electron microscope
image of the sample with edge channels. The gates defining the QPC
are coloured yellow. A source contact is placed at bias voltage $V$,
emitting a chirally propagating current, shown as a red line. Upon
arriving at the QPC, the current is partitioned, with parts propagating
along the respective edge channels on the two sides of the QPC (red-white
dashed lines). An amplifier (bottom-right of the image) measures the
excess auto-correlation noise $S$ and the differential conductance
$\tilde{g}=-dI/dV+\nu e^{2}/h$.}
\label{fig:system}
\end{figure}

One should note that recently CLL behaviour has been convincingly
demonstrated in a graphene-based FQH system \citep{Cohen2023}. These
results are closely related to the present paper and are discussed
towards its end (see Sec.~\ref{subsec:discussion_tun_curr_experiments}).

Measuring both tunnelling current and its noise is customarily considered
a more reliable method to probe the QPC physics. The most important
quantity in such experiments is the Fano factor, $F=S/(2eI)$. Here
$S$ denotes the low-frequency spectral density of noise, while $I$
is the tunnelling current, and $e$ is the electron charge. In the
shot noise limit, $eV\gg k_{B}T$ ($T$ is the system temperature),
the Fano factor is predicted to give the quasiparticle charge, $q$
(in units of the electron charge). Noise measurements have been used
to extract the fractional charges at the filling factor of $\nu=1/3$
and a myriad other filling factors \citep{de-picciotto_direct_1997,saminadayar_observation_1997,griffiths_evolution_2000,dolev_observation_2008}.

Some experiments reported unexpected behaviours such as the extracted
charge changing with the system temperature \citep{Chung2003} or
bias voltage \citep{dolev_observation_2008}, or other parameters
affecting the system \citep{Bid2010,Snizhko2016}. Even for $\nu=1/3$,
deviation from the predicted $q=e/3$ behaviour can be seen at larger
bias voltages in the experimental data \citep{de-picciotto_direct_1997,saminadayar_observation_1997}.
Moreover, recently it has been observed that in the presence of neutral
modes the Fano factor may correspond to the bulk filling factor, and
not to the quasiparticle charge \citep{biswas_does_2021}. This suggests
that Fano-factor-based experiments cannot be tasked with determining
the specific value of the fractional charge without additional input
or assumptions.

Note, however, that the presence of $e/3$-charged quasiparticles
in the $\nu=1/3$ Laughlin state has been corroborated by multiple
works based on drastically different experimental methods in unrelated
material samples \citep{Goldman1995,Martin2004,Kapfer2019,Bisognin2019a}.
Therefore, the charge of the elementary quasiparticle at $\nu=1/3$
can be taken as $q=e/3$ with a high value of certainty.

The above literature review shows that while there are numerous signatures
of agreement in the basic predictions of the CLL models and the experimental
observations, rarely do the theory and experiment demonstrate accurate
agreement without ad-hoc modifications or (often implicit) assumptions.
A priori, these discrepancies may be attributed to either complete
inadequacy of the standard theory framework or to non-idealities which
cause deviations from the ideal CLL behaviour. \emph{It is, thus,
important to minimize the influence of non-idealities and check the
CLL framework as such.}

The Fano factor in the regime of weak quasiparticle tunnelling turns
out to be an ideal candidate for performing such a check. One can
write the universal answer for \emph{any} model within the CLL framework
\citep{schiller_extracting_2022,snizhko_scaling_2015,shtanko_nonequilibrium_2014}:
\begin{equation}
F=\frac{2}{\pi}q\,\mathrm{Im}\left[\psi\left(2\delta+i\frac{qeV}{2\pi k_{B}T}\right)\right],\label{eq:digamma}
\end{equation}
where $\mathrm{Im}\left[\cdots\right]$ denotes the imaginary part,
and $\psi$ is the \emph{digamma function}. Equation~\eqref{eq:digamma}
recreates the asymptotic shot noise limit, $F\rightarrow q\,V/\abs V=q\,\sgn\,V$
for $e\left|V\right|\gg k_{B}T$. Yet, when $e\left|V\right|$ and
$k_{B}T$ are comparable, the formula enables extracting the scaling
dimension $\delta$ alongside the quasiparticle charge $q$. There
exist non-idealities that lead to a renormalization of the scaling
dimension (such as the ones mentioned above) --- yet they do not
change Eq.~(\ref{eq:digamma}) unless they break the CLL. Further,
Eq.~(\ref{eq:digamma}) is not sensitive to the tunnelling amplitude
at the QPC, which may exhibit non-universal dependence on the system
parameters (such as $V$ or $T$).

\emph{The most important feature of Eq.~(\ref{eq:digamma}) for the
present work is the }\textbf{\emph{scaling behaviour}}\emph{:} The
Fano factor depends only on the ratio $x\equiv eV/(2\pi k_{B}T)$,
and not on $eV$ or $k_{B}T$ separately. Therefore, the Fano factor
data measured at different temperatures should collapse on top of
each other when plotted as a function of this dimensionless ratio.
In fact, this prediction can be expected even without reference to
the CLL: as long as $eV$ and $k_{B}T$ are the \emph{only} relevant
energy scales, the dimensionless Fano factor can only depend on their
dimensionless ratio.

Conversely, the collapse may not be observed in the presence of other
energy scales. The bulk gap is one such energy scale, yet other scales
may exist.

In this work, we perform experiments at $\nu=1/3$. We measure the
noise and the differential conductance in the QPC tunnelling processes
at several temperatures, and then analyze the data using Eq.~(\ref{eq:digamma}).
We first verify the \emph{scaling behaviour} of the Fano factor. We
find a partial data collapse, hinting at the violation of CLL behaviour.
The region where the scaling behaviour does hold, corresponds to $x\equiv eV/(2\pi k_{B}T)\lesssim1.5\text{--}2.5$.
The data in this region only enable extracting a combination of $q$
and $\delta$, and not each of them separately. Assuming the value
of $q=e/3$, as \emph{known from experiments of other types} \citep{Goldman1995,Martin2004,Kapfer2019,Bisognin2019a},
we extract the value of $\delta\approx1/2$. This value is drastically
different from the naïve CLL prediction of $\delta=1/6$ for $\nu=1/3$.

Our results suggest the existence of physical effects leading to scaling
dimension renormalization, and to generation of additional energy
scales which lead to the CLL violation significantly below the bulk
gap. We identify electrostatic interactions at the QPC as a likely
culprit.

Most importantly, the present firm observation of the breakdown of
the CLL and of the scaling dimension renormalization in the CLL-compatible
regime opens a way to reconcile various experimental results on QPC
tunnelling with the CLL theory of fractional quantum Hall edges \emph{in
a systematic manner}. Ad-hoc modifications to the theory can now be
replaced with separating CLL-compatible and CLL-incompatible regions
of data. Furthermore, our method opens a new window into the mesoscopic
world physics by allowing \emph{quantitative} studies of scaling dimension
renormalization.

\section{Results}

We perform standard measurements of the tunnelling current and noise
generated due to quasiparticle tunnelling at $\nu=1/3$, as shown
in Fig.~\ref{fig:system}. The source contact is placed at bias voltage
$V_{S}\equiv V$, emitting a chirally propagating current, shown as
a red line. This current arrives at a QPC, where it is partitioned.
An amplifier is connected to the Ohmic contact downstream from the
QPC, enabling the measurement of the excess\footnote{``Excess'' means the noise with the Johnson-Nyquist noise (the noise
measured at $V=0$) subtracted.} auto-correlation noise $S$ and the differential conductance $\tilde{g}=\nu e^{2}/h-g$,
where $I$ is the tunnelling current leaking from the edge via the
QPC and $g=dI/dV$ is the differential tunnelling conductance. The
Fano factor, $F=S/(2eI)$, is calculated based on these data. All
the measurements have been performed on a 2D electron gas embedded
in a GaAs/AlGaAs heterostructure, with an areal density of $n=9.7\times10^{10}\text{ cm}^{-2}$
and mobility $\mu=4.0\times10^{6}\text{ cm}^{2}\mathrm{V}^{-1}\mathrm{s}^{-1}$
at 4~K temperature, on which the QPC was fabricated. Additional data
on the sample, fabrication and measurement is given in the Methods
section.

\begin{figure*}
\centering{}\includegraphics[width=1\textwidth]{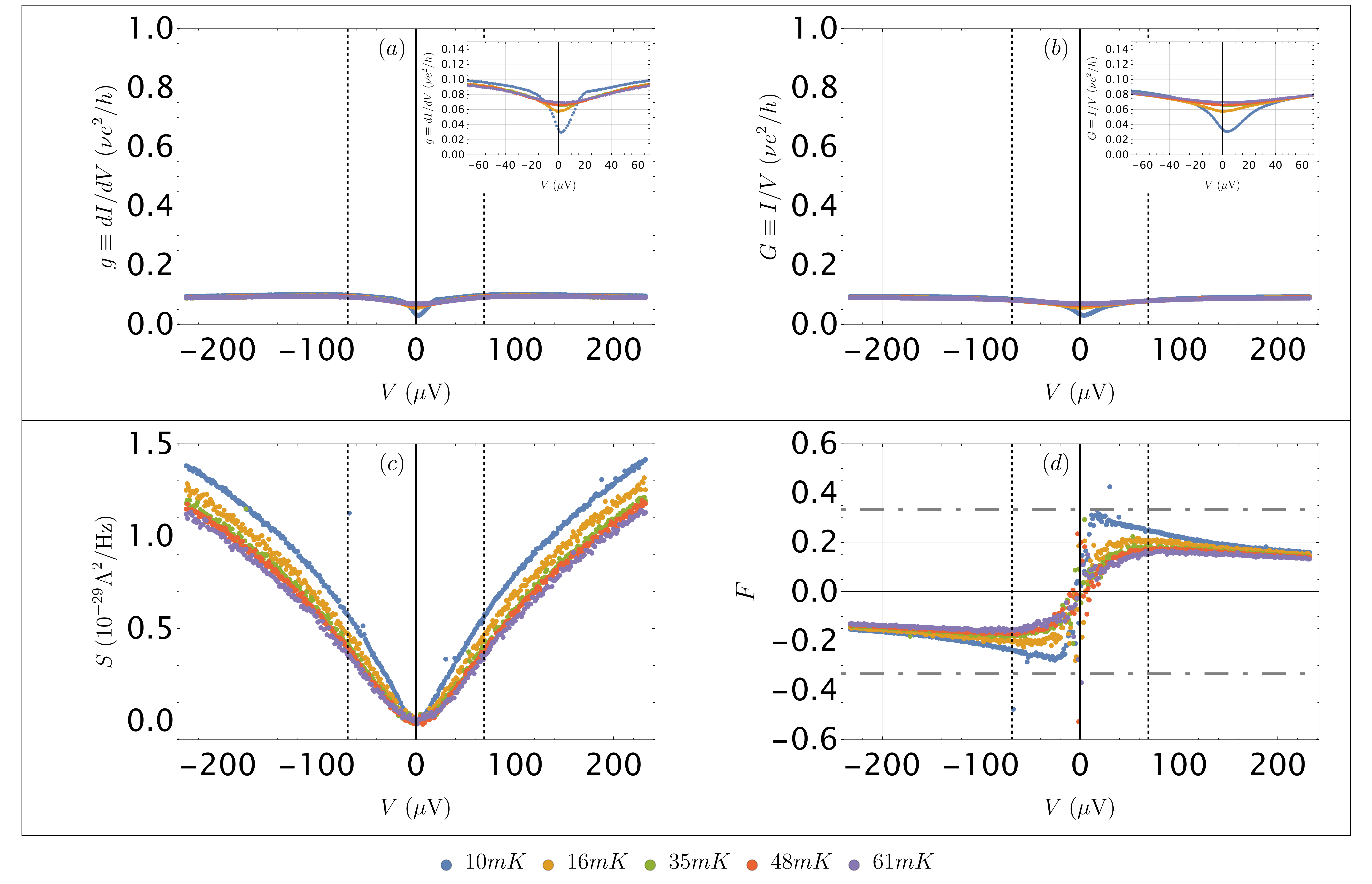} \caption{The tunnelling data measured at five sample temperatures. (a) ---
the measured differential conductance $g=dI/dV$. (b) --- the integrated
conductance $G=I/V$ (calculated from $g$). (c) --- excess noise
$S$. (d) --- Fano factor $F=S/(2eI)$ (calculated from $G$ and
$S$). The change of sign of the Fano factor is due to the current
$I$ being negative for negative voltages. The vertical dashed lines
in all panels correspond to $eV=\pm E_{\mathrm{gap}}/5$, where the
bulk gap energy $E_{\mathrm{gap}}$ for $\nu=1/3$ state is conservatively
taken as $4\,\mathrm{K}\approx345\,\mu\mathrm{eV}$, cf.~Refs.~\citep{Pan2020c,Rosales2021}.
The insets in panels (a) and (b) zoom in on the data from the main
plots. In panel (d), the grey dot-dashed lines correspond to $F=\pm1/3$,
the expected value of $F$ at $e\left|V\right|\gg k_{B}T$. The sample
temperatures $10$--$61\,\mathrm{mK}$ correspond to $V=2\pi k_{B}T/e\approx5$--$35\,\mathrm{\mu V}$
well within the data range. Yet the data clearly and systematically
deviate from the expected value $F=q\,\protect\sgn\,V=\pm1/3$, indicating
inconsistency with the conventional expectation for the shot noise.
At the same time, the energy scale $eV$ is significantly below $E_{\mathrm{gap}}$,
making it unlikely that the bulk transport plays a significant role.
Note that such behaviour at large $V$ has been observed even in the
earliest Fano-factor-based measurements of the quasiparticle charge
$q$, cf.~\citep[Figs. 3, 4 in][]{de-picciotto_direct_1997} and
\citep[Figs. 2, 3 in][]{saminadayar_observation_1997}. Such behaviour
is thus not specific to the present experiment, but is rather widespread.}
\label{fig:raw_data}
\end{figure*}

The measurements are performed at five different temperatures of the
sample. The electron temperature has been inferred from the Johnson-Nyquist
noise measurement (i.e., at vanishing bias voltage, $V=0$). The raw
data for the differential conductance $g=dI/dV=\nu e^{2}/h-\tilde{g}$
and the excess noise $S$ are presented in Fig.~\ref{fig:raw_data}(a,c).
Note that for all data points $g\lesssim0.1\,\nu e^{2}/h$, indicating
a regime of weak quasiparticle tunnelling, in which Eq.~(\ref{eq:digamma})
is expected to be valid. Based on the raw data, the integral conductance
$G=I/V$ and the Fano factor $F=S/(2eI)$ are calculated, see Fig.~\ref{fig:raw_data}(b,d).

Note that the theoretical prediction for the Fano factor in the regime
of large $V$, $F=q\,\sgn\,V=\pm1/3$, is in stark contrast with the
experimental observations. This discrepancy was observed even in the
earliest Fano-factor-based measurements of the fractional charge,
cf.~\citep[Figs. 3, 4 in][]{de-picciotto_direct_1997} and \citep[Figs. 2, 3 in][]{saminadayar_observation_1997}.
We will come back to this issue in the Discussion section. For the
moment, we only remark that the raw experimental data do not contain
a regime of the Fano factor leveling off at $F=q\,\sgn\,V=\pm1/3$
for an extended range of voltages, in contradiction with the conventional
expectation by theorists.

\begin{figure}
\begin{centering}
\includegraphics[width=1\columnwidth]{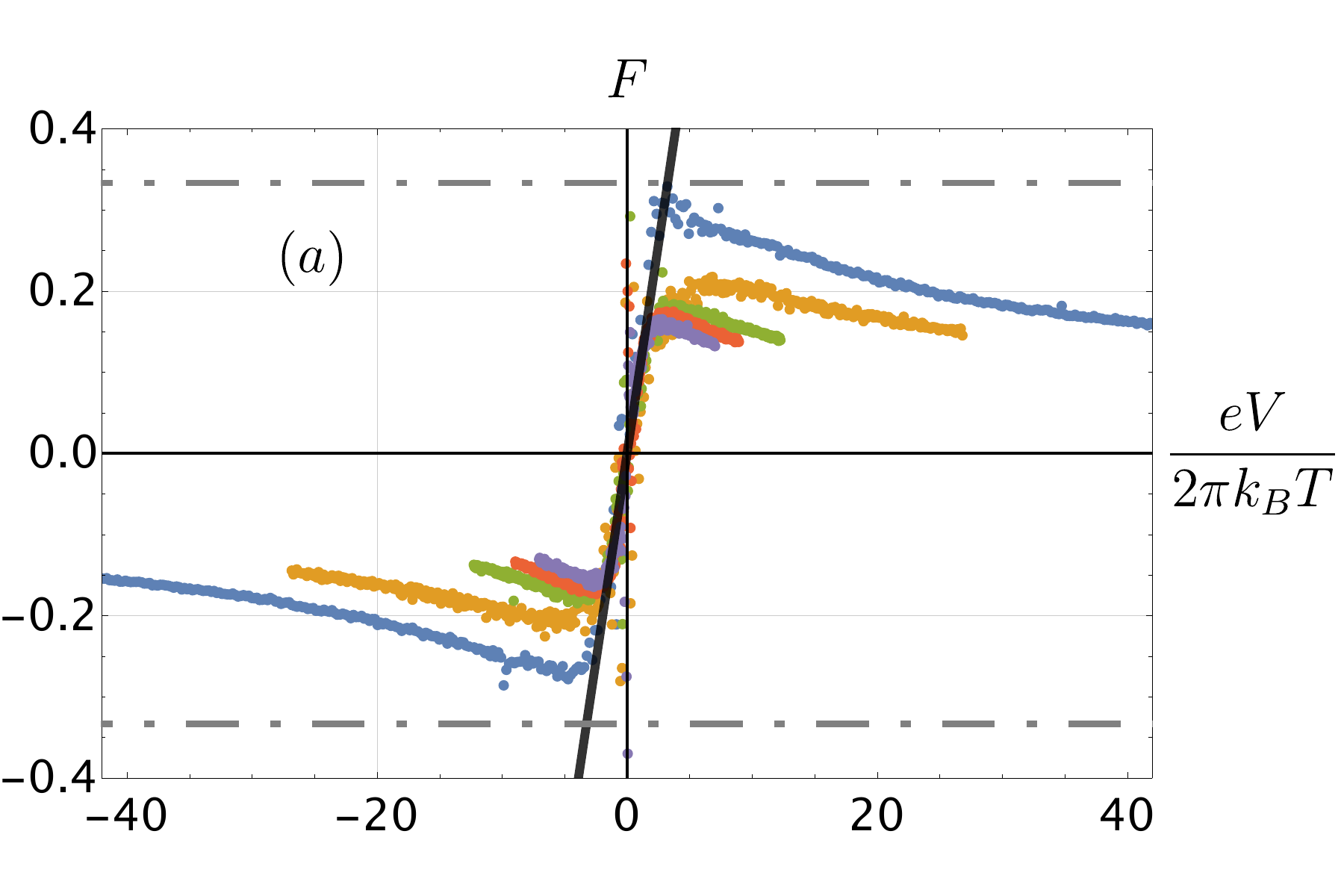} 
\par\end{centering}
\begin{centering}
\includegraphics[width=1\columnwidth]{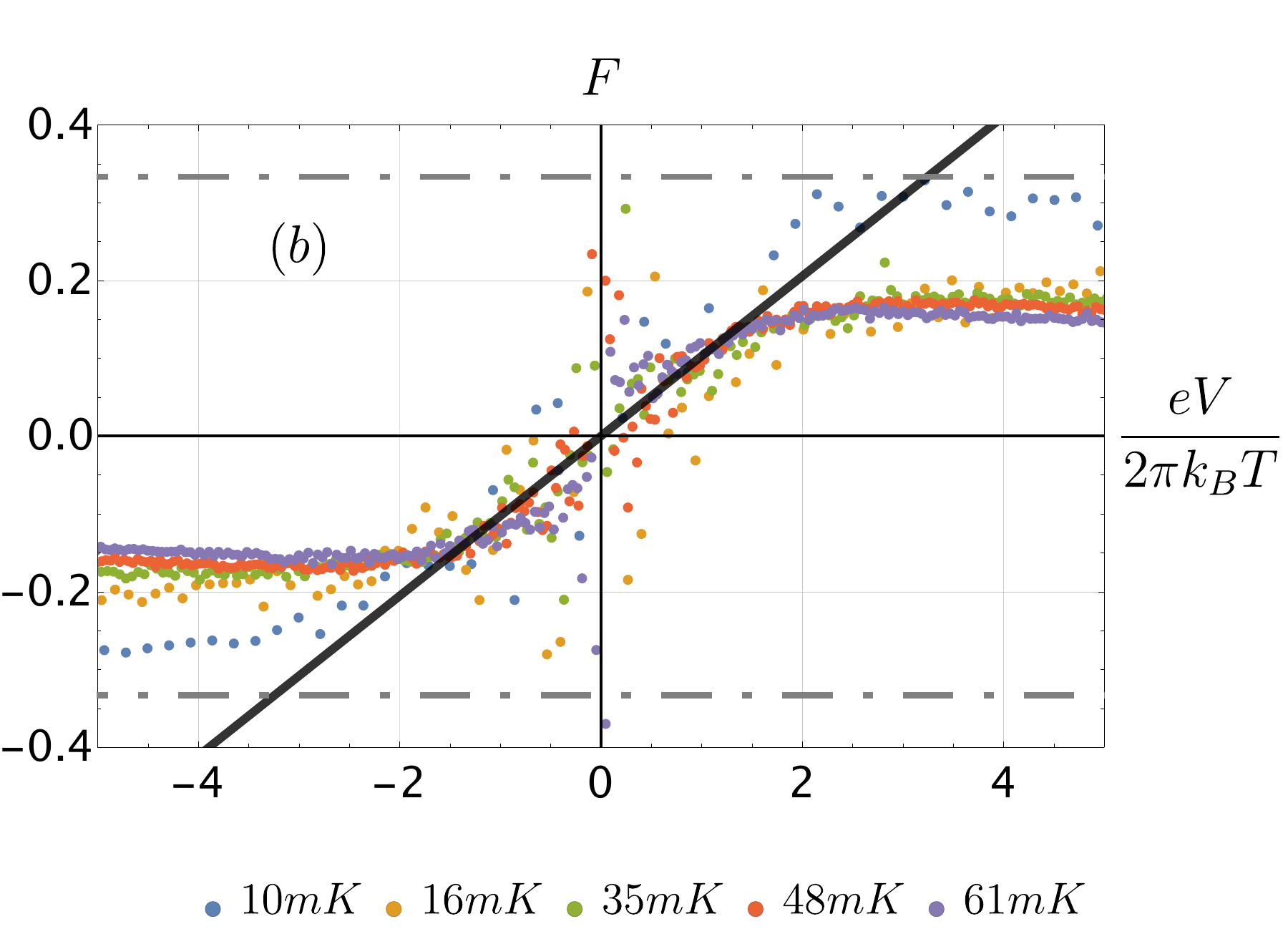} 
\par\end{centering}
\centering{}\caption{Measured Fano factors as a function of the unitless variable $x=eV/2\pi k_{B}T$.
Five temperatures were measured. Scaling can be seen solely in the
low voltage range, $x\lesssim2$, where Eq.~\eqref{eq:digamma} predicts
linear behaviour (\ref{eq:digamma_linear}). The high voltage behaviour
falls short of the predicted $F=\pm q$ asymptotes (grey dot-dashed
lines) for all temperatures. A linear fit (solid black line) corresponds
to the fitted value of $\frac{2}{\pi}q^{2}\psi^{\prime}(2\delta)=0.103\pm0.012$.
(a) --- full range of data. (b) --- zoom-in on low voltage range,
where scaling is observed.}
\label{fig:scaled_Fano}
\end{figure}

As stated in the introduction, the primary model-independent test
for CLL behaviour is the scaling property of the Fano factor. The
scaled data for the Fano factor are presented in Fig.~\ref{fig:scaled_Fano}.
We observe the data collapse for $\abs{eV/(2\pi k_{B}T)}\lesssim1.5$,
whereas beyond that region the scaling fails. The three higher-temperature
curves seem to collapse well up to a higher cutoff of $\abs{eV/(2\pi k_{B}T)}\lesssim2.5$,
yet also part from each other outside that region. These deviations
from the expected scaling behaviour hint at a violation of CLL behaviour
at an energy scale of the order $eV_{\mathrm{breakdown}}=2\pi k_{B}Tx^{*}=0.1\text{--}1\text{ K}=8\text{--}80\text{\,\ensuremath{\mu}eV}$.
Here $x^{*}$ is the breakdown point of the scaling; the $0.1$~K
estimate is produced using $x^{*}=1.5$ and $T=10$~mK; the $1$~K
estimate follows from taking $x^{*}=2.5$ and $T=61$~mK. Note that
even $1$~K is significantly below the bulk gap scale $E_{\mathrm{gap}}\gtrsim4$~K
(the estimate is taken as the smallest gap for $\nu=1/3$ measured
in Refs.~\citep{Pan2020c,Rosales2021}). One is forced to conclude
that an important additional energy scale (perhaps, several) is present
below the bulk gap.

\begin{figure}
\centering{}\includegraphics[width=1\columnwidth]{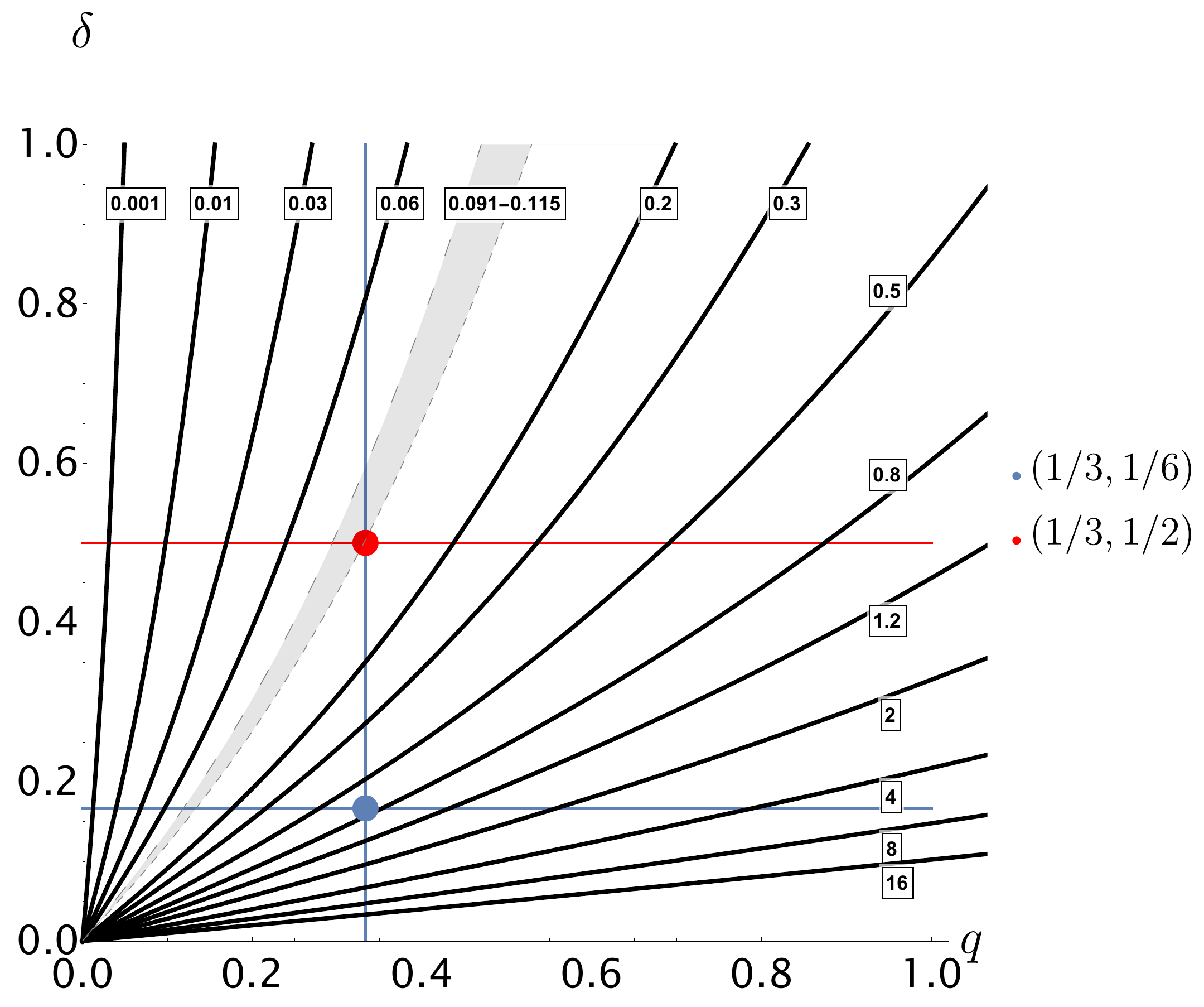}
\caption{Equal value contours of the function $\frac{2}{\pi}q^{2}\psi^{\prime}(2\delta)$
in the $q\text{--}\delta$ plane. The shaded grey region represents
the upper and lower bounds of the value obtained from Fig.~\ref{fig:scaled_Fano}.
The naïve theoretical point $q=1/3,\delta=1/6$ is inconsistent with
the measurement. Rather, the measurement crosses the $q=1/3$ line
(blue vertical) at $\delta\approx1/2$, and the $\delta=1/6$ line
(blue horizontal) at $q\approx0.13$.}
\label{fig:contours} 
\end{figure}

Henceforth, we focus on the region of sufficiently small $x$, where
scaling behaviour is observed and which can thus be compatible with
the CLL. The collapsed data are linear in $x=eV/(2\pi k_{B}T)$, which
corresponds to the regime of $e\abs V\ll k_{B}T$ in Eq.~(\ref{eq:digamma}),
where it can be approximated as 
\begin{equation}
F=\frac{2}{\pi}q^{2}\psi^{\prime}(2\delta)x.\label{eq:digamma_linear}
\end{equation}
This implies that the data in this regime do not allow for extracting
the quasiparticle charge $q$ and scaling dimension $\delta$ individually,
only their combination. We fit the data and extract the slope $\frac{2}{\pi}q^{2}\psi^{\prime}(2\delta)=0.102\pm0.015$
(95\% confidence). Figure~\ref{fig:contours} displays contours for
different values of $\frac{2}{\pi}q^{2}\psi^{\prime}(2\delta)$ in
the $q\text{--}\delta$ plane; the region compatible with the extracted
slope is highlighed in grey. Notably, the pristine theoretical prediction
for the Laughlin edge, $q=1/3,\delta=1/6$, is incompatible with this
curve. Assuming that the quasiparticle charge is $q=1/3$ (as evidenced
by multiple experiments \citep{Goldman1995,Martin2004,Kapfer2019,Bisognin2019a}),
one extracts $\delta=0.54\pm0.05$.

It is worth elaborating further on the particular value of $\delta\approx1/2$
extracted above. The CLL predicts the tunnelling current to behave
as\textcolor{red}{{} }
\begin{equation}
I=\frac{\zeta}{\pi}(2\pi k_{B}T)^{4\delta-1}\sinh\left(\frac{qeV}{2k_{B}T}\right)\abs{\Gamma\left(2\delta+i\frac{qeV}{2\pi k_{B}T}\right)}^{2},\label{eq:current_CLL}
\end{equation}
where $\Gamma\left(x\right)$ is the Euler's gamma function and $\zeta$
encodes the bare tunnelling amplitude for quasiparticles at the QPC
(see Refs.~\citep{Wen1991,kane_transport_1992,kane_transmission_1992,Chamon1993,kane_nonequilibrium_1994}
for the original derivation and Refs.~\citep{Chamon1997,Roddaro2003,snizhko_scaling_2015}
for the expression in the notation close to the present one). In the
regime $eV\gg k_{B}T$, this reduces to 
\begin{equation}
I=\zeta\left(qeV\right)^{4\delta-1}.\label{eq:current_asymp}
\end{equation}
The differential conductance $g=dI/dV\propto V^{4\delta-2}$ is predicted
to exhibit a zero-bias peak for $\delta<1/2$ and a zero-bias dip
for $\delta>1/2$. The \emph{value} of $\delta=1/2$ is associated
with $g=\mathrm{const}$. This is compatible with the behaviour seen
in Fig.~\ref{fig:raw_data}(a) (in fact, the zero bias dip at small
voltage is compatible with the value slightly above 1/2, which we
have extracted).

This correspondence between the scaling dimension and the differential
conductance could, in principle, be spoiled. The tunnelling amplitude
could have a dependence of the bias voltage $V$, due to electrostatic
interactions in the sample forcing the QPC edges to change their location
depending on the bias voltage. This would transform Eq.~(\ref{eq:current_asymp})
to be $I=\zeta(V)\left(qeV\right)^{4\delta-1}$, and the differential
conductance $g$ would acquire an additional non-universal contribution
$\propto d\zeta(V)/dV$, disallowing one to make conclusions about
the scaling dimension from the tunnelling conductance alone. The fact
that the scaling dimension extracted from the Fano factor also enables
explaining the behaviour of $g$ is thus an evidence that $\zeta(V)$
dependence can be neglected. If one \emph{assumes} that such a statement
has general applicability, one obtains additional means of verifying
the value of $\delta$ independently.

\section{Discussion}

The above results provide clear evidence for heavy renormalization
of the scaling dimension and provide an experimental method for its
\emph{quantitative} investigation. However, the implication of these
results is much more significant, as it enables reconciliation of
multiple experimental results and theoretical efforts of the last
25+ years. Below, we discuss various aspects of those links and interconnections.

\subsection{\label{subsec:discussion_tun_curr_experiments}Relation to other
tunnelling current experiments}

The obtained value of $\delta\approx1/2$ may seem quite arbitrary.
However, the appearance of this specific value has deep ties to how
Fano factor experiments have been performed. In a typical experimental
sample, the observed dependences $g(V)$ of the differential conductance
on the bias voltage change from zero-bias peaks to zero-bias dips
as a function of the voltage of the gates that confine the QPC, see,
e.g., \citep[Figs. 3, 5 in][]{Heiblum2006} or \citep[Fig. S3c in the SM of][]{bartolomei_fractional_2020}.
With the few exceptions we will discuss below, past experiments have
focused on the regime of as flat $g(V)$ as possible. This practical
choice has two reasons. First, the CLL predictions are well controlled
in the regime when the tunnelling conductance is small ($g\ll\nu e^{2}/h$);
satisfying this requirement for the whole range of $V$ is easiest
when the dependence $g(V)$ is flat. Second, this choice is influenced
by the intuition coming from dealing with the Fermi liquid; non-interacting
free fermions can be described with CLL using $\delta=1/2$, again
predicting flat $g(V)$; with such intuition, deviation from this
behaviour may raise suspicions of non-universal behaviour of the tunnelling
amplitude $\zeta(V)$ at the QPC --- a good thing to avoid, when
trying to verify universal predictions.

Therefore, obtaining $\delta=1/2$ appears to be a consequence of
deliberate preference for using the data with flat $g(V)$. \emph{One
implication of the above is that the data with non-flat transmission,
which are typically ignored, can be analyzed with the help of our
method.} This analysis would allow for a systematic study of the scaling
dimension renormalization in the QPCs.

While experiments aiming at extracting the Fano factor have historically
focused on flat $g(V)$, there have been a number of experiments aiming
to measure the CLL-predicted zero-bias peak, \citep[Fig. 3c in][]{Heiblum2006}
and \citep{Roddaro2003,Roddaro2004,radu_quasi-particle_2008,Lin2012,rossler_experimental_2014}.
These experiments used the data in the regime where $g(V)$ exhibits
a zero-bias peak. They found good agreement between the theory prediction
(\ref{eq:current_CLL}) and the experimental data at sufficiently
small bias voltages $V$, yet deviations from the predicted behaviour
at larger $V$. The results of our analysis suggest that these deviations
are due to the violation of CLL behaviour we find at sufficiently
large energies.

Further, some experiments hunting for the zero-bias peak \citep{radu_quasi-particle_2008,Lin2012,rossler_experimental_2014}
have made ad-hoc adjustments to the theory by adding an extra constant
to $g(V)$, in order to account for flat transmission at high $V$.
This enabled extracting the values of $q$ and $\delta$ by fitting
the experimental data with modified formulas. However, the extracted
values never fully coincide with the theory predictions for $(q,\delta)$.
Our results suggest a better way to analyse the data: cut out the
region of $V$ where CLL is no longer valid (where scaling of the
Fano factor fails) and analyze the data without making ad-hoc adjustments.
Indeed, we show in the Supplemental Material \citep{SupplMat} that
ignoring the CLL breakdown can lead to data fits of excellent quality,
yet with wrong and even temperature-dependent values of $(q,\delta)$.

Finding the appropriate region of $V$ to analyze constitutes a challenge
when data only at one temperature are available. Having the data at
several temperatures, however, makes this possible.

A recent work \citep{Cohen2023} has pointed out that CLL predicts
scaling behaviour for $g(V,T)$ and verified this behaviour in a FQH
QPC in graphene.\footnote{Admittedly, such behaviour was demonstrated in a rather exotic situation:
electron tunnelling between the edges of different filling factors.
Note, however, that such results have not been demonstrated so far
for quasiparticle or electron tunnelling between the edges of the
same filling factor.} This method assumes that the tunnelling amplitude $\zeta$ does not
exhibit non-universal dependence $\zeta(V)$. Under this assumption,
this method enables determining the energy scale of CLL violation
similarly to ours. Combining the two methods, would allow a powerful
experimental methodology with internal cross-checks. In the Supplementary
material \citep{SupplMat}, we demonstrate that this method works
with our data for all temperatures except the lowest one. This may
indicate the importance of excluding the influence of $\zeta(V)$
via our Fano factor scaling technique.

\subsection{\label{subsec:discussion_sc_dim_renormalization}Scaling dimension
renormalization}

The scaling dimension we find, $\delta\approx0.54\pm0.05$, is clearly
distinct from the naïve CLL prediction of $\delta=1/6$ for the Laughlin
edge at $\nu=1/3$. This on its own, however, does not invalidate
the CLL framework for describing quantum Hall edges. Over the years,
theorists proposed a number of effects that can lead to scaling dimension
renormalization. First and foremost, the same filling factor admits
multiple edge models \citep{wen_quantum_2007} that can have a different
number of edge modes, different spectrum of quasiparticles, and the
elementary quasiparticle in those models can differ in charge, statistics,
and scaling dimension. Further, in the presence of counterpropagating
edge modes, the scaling dimension of a quasiparticle can change due
to electrostatic interactions between those edge modes \citep{KaneFisherPolchinski,KaneFisher,Rosenow2002}.
Finally, even the simplest edges allow for scaling dimension renormalization
locally, in the vicinity of a QPC due to interactions across the elements
of the QPC \citep{Pryadko2000,papa_interactions_2004,yang_influence_2013}.

The continuous change of shape of the differential conductance curves
as a function of the QPC confining voltage, see \citep[Figs. 3, 5 in][]{Heiblum2006}
and \citep[Fig. S3c in the SM of][]{bartolomei_fractional_2020},
suggests that the latter mechanism plays the dominant role in QPC
experiments. The method proposed above enables quantitative experimental
studies of scaling dimension renormalization. 

Back-of-the envelope estimates for CLL breakdown scale also speak
in favour of this mechanism. In our sample, the distance $L_{\mathrm{gate}}$
along which the incoming and outgoing parts of the edge at the QPC
are close to each other is on the order of $1\text{\,\ensuremath{\mu}m}$,
cf.~Fig.~\ref{fig:system}. This is the region where one can expect
the Coulomb interaction between the edges to be strong, which renormalizes
the scaling dimension. The pattern of renormalization should, however,
depend on the energies concerned \citep{yang_influence_2013,zucker_edge_2015}:
for energies up to $E_{\text{typ}}=\hbar v/L_{\mathrm{gate}}$, the
wavelength of incoming quasiparticles is larger than $L_{\mathrm{gate}}$,
while for $E>E_{\text{typ}}$, the wavelength fits completely within
the region of strong Coulomb interactions. One, therefore, expects
a change of renormalization pattern and thus CLL breaking around $E_{\text{typ}}$.
Taking $v\sim2\times10^{4}\text{\,m/s}$ \citep{Gurman2016} and $L_{\mathrm{gate}}\sim1\text{\,\ensuremath{\mu}m}$,
one estimates $E_{\text{typ}}=\hbar v/L_{\mathrm{gate}}\sim0.15\text{\,K}=13\text{\,\ensuremath{\mu}eV}$.
This order-of-magnitude estimate agrees surprisingly well with the
CLL breakdown scale ($eV_{\mathrm{breakdown}}=2\pi k_{B}Tx^{*}=0.1\text{--}1\text{ K}=8\text{--}80\text{\,\ensuremath{\mu}eV}$)
extracted from the experimental data.

As for detailed theory predictions of the renormalization, it is possible
to predict the scaling dimension using CLL if one knows the strength
of electrostatic interactions across the QPC \citep{Pryadko2000,papa_interactions_2004,yang_influence_2013}.
Coincidentally, numerical methods developed recently allow for quantitatively
accurate predictions of the electrostatic behaviour of mesoscopic
devices \citep{Flor2022,Chatzikyriakou2022}. Combining these analytical
and numerical methods together, it should be possible to derive predictions
for the QPC behaviour based solely on the knowledge of the material
properties and device geometry.

\subsection{\label{subsec:discussion_earlier_qp_charges}Interpreting the quasiparticle
charges reported previously based on the Fano factor}

As stated above, the theoretical expectation that the Fano factor
at sufficiently large bias voltage $V$ becomes the quasiparticle
charge, $F=q$, does not hold in actual experiment. This raises the
question: what should one think of the previously reported fractional
charges extracted from the Fano factor experiments at various filling
factors?

\begin{figure}
\centering{}\includegraphics[width=1\columnwidth]{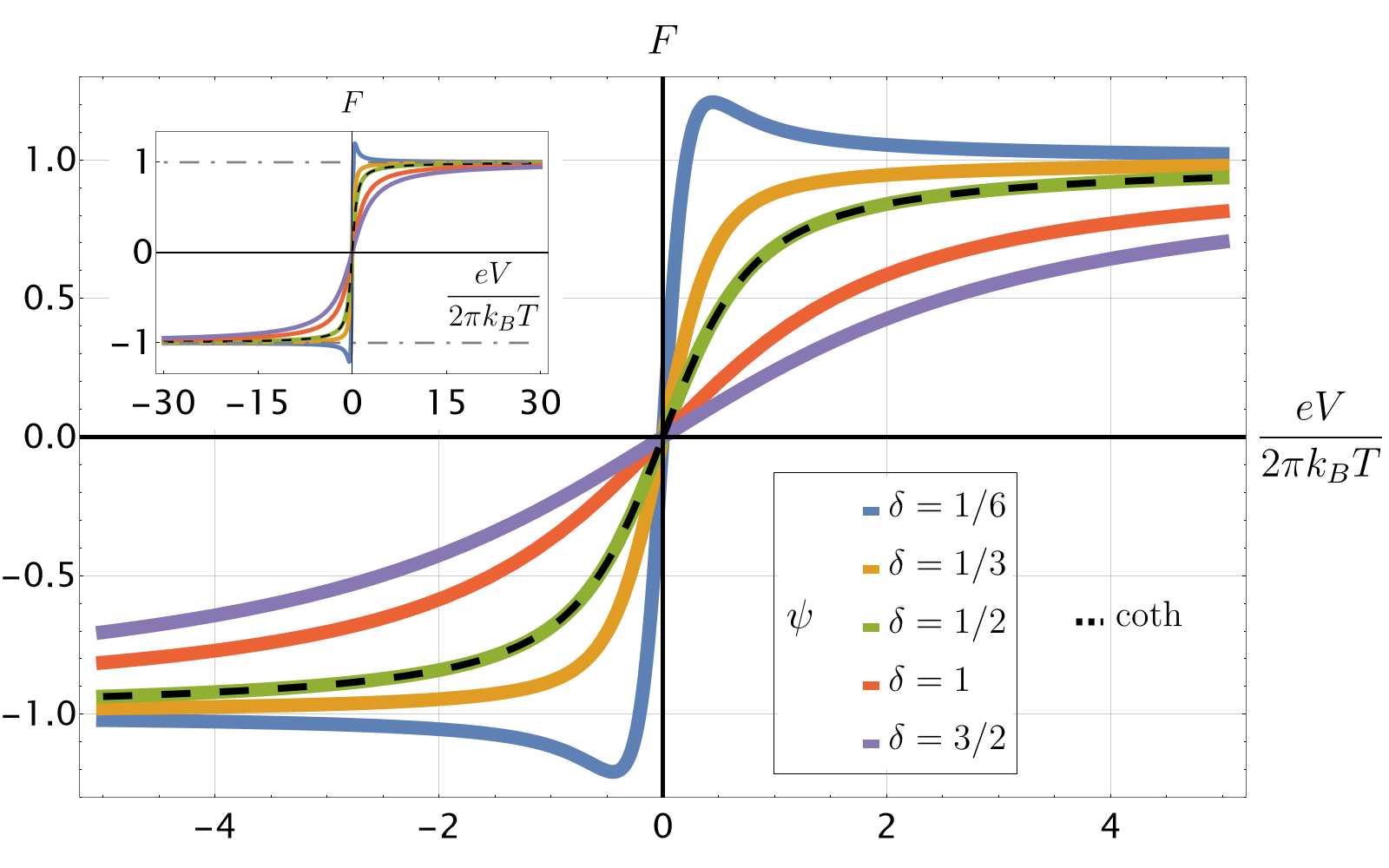}
\caption{Theoretical Fano factors as a function of the unitless variable $x\equiv eV/(2\pi k_{B}T)$,
for charge $q=1$. The colorful lines represent Fano factor dependences
predicted by the CLL, Eq.~\eqref{eq:digamma}, for different values
of $\delta$ (\textquotedblleft digamma\textquotedblright , $\psi$).
The dashed black line represents the Fano factor of the phenomenological,
non-interacting theory, Eq.~\eqref{eq:Fano_coth} for $g\ll\nu e^{2}/h$
(\textquotedblleft$\coth$\textquotedblright ). Note that for $\delta=1/2$
the CLL and the phenomenological $\coth$ theory yield the same prediction.
At the same time, the behaviour for $\delta\protect\neq1/2$ is distinct
and distinguishable (cf.~Fig.~\ref{fig:contours}) from the $\coth$
prediction. Further, for $\delta<1/4$ (represented by $\delta=1/6$
in the plot), the Fano factor dependence on the voltage is predicted
by the CLL to be non-monotonous --- in stark contrast with the coth
prediction. Inset: all curves in both theories asymptotically approach
$F=\pm q$ at $x\rightarrow\pm\infty$.}
\label{fig:digamma_v_coth}
\end{figure}

Providing an answer to the question requires a close look at the methodology
that was used to extract those values of $q$. The conventional methodology
is as follows \citep[e.g., ][]{Dolev2010,biswas_does_2021}: tune
the QPC to the regime of flattest possible differential conductance
dependence $g(V)$ and analyze the experimental data using a phenomenological
Fermi-liquid-inspired formula,
\begin{equation}
F=q\left(1-g\frac{h}{\nu e^{2}}\right)\left[\coth\left(\frac{qeV}{2k_{B}T}\right)-\frac{2k_{B}T}{qeV}\right].\label{eq:Fano_coth}
\end{equation}
This formula has the expected behaviour of $F=q\,\sgn\,V$ when $e\abs V\gg k_{B}T$
and $g\ll\nu e^{2}/h$. The formula does not have any dependence on
the scaling dimension $\delta$, making it less general than Eq.~(\ref{eq:digamma}),
cf.~Fig.~\ref{fig:digamma_v_coth}. However, for $\delta=1/2$ Eq.~(\ref{eq:digamma})
reduces to Eq.~(\ref{eq:Fano_coth}) taken in the limit $g\ll\nu e^{2}/h$.
Note that $\delta=1/2$ corresponds to free non-interacting fermions,
but it also describes fractional quasiparticles whose scaling dimension
was renormalised to the value of $1/2$. Given the results of our
analysis above, it stands to reason that tuning $g(V)$ to be flat
corresponds to tuning $\delta$ to be $1/2$. Therefore, by serendipity,
the use of Eq.~(\ref{eq:Fano_coth}) in this context is justified
and expected to return the correct value of $q$.

Concerning the unexpected large-$V$ behaviour of the Fano factor,
such behaviour has also been observed in the past. See, e.g., Refs.~\citep[Figs. 3, 4 in][]{de-picciotto_direct_1997}
and \citep[Figs. 2, 3 in][]{saminadayar_observation_1997} --- it
is widespread, and not specific to the present experiment. The only
behaviour of the Fano factor allowed by Eq.~(\ref{eq:Fano_coth})
is for the Fano factor to level off after the initial period of growth,
cf.~Fig.~\ref{fig:digamma_v_coth}. Therefore, the Fano factor bending
down as in Fig.~\ref{fig:raw_data}(c) was a reason to believe that
the theory has stopped working at this energy scale and the data for
a limited range of $V$ should be analysed. Coincidentally, this happens
around the same place where the scaling behaviour breaks down. Given
the tuning to $\delta=1/2$ discussed above, this analysis of the
truncated data --- using Eq.~(\ref{eq:Fano_coth}) --- leads to
the same results as in our more general methodology. At the same time,
we stress that for $\delta<1/4$ this intuition would be misleading
and could lead to discarding valid data in the CLL regime, cf.~Fig.~\ref{fig:digamma_v_coth}.

Interestingly, there have been theoretical attempts to justify the
use of Eq.~(\ref{eq:Fano_coth}) or similar coth-based formulas for
noise analysis \citep{trauzettel_effect_2004,feldman_why_2017}. We
believe that these theory arguments are correct, but require careful
interpretation. First, we note that the predictions of Eq.~(\ref{eq:Fano_coth})
are clearly distinguishable from those of Eq.~\eqref{eq:digamma}
in principle (as evidenced by Fig.~\ref{fig:digamma_v_coth}) and
in practice. Indeed, Fig.~\ref{fig:contours} shows that the naive
CLL prediction of $(q=1/3,\delta=1/6)$ is clearly excluded by the
experimental data, while $(q=1/3,\delta\approx1/2)$ is not.

Second, the two references \citep{trauzettel_effect_2004,feldman_why_2017}
explicitly have $g(V)$ in their coth-based formulas\footnote{For example, through using Eq.~\eqref{eq:digamma} outside the $g\ll\nu e^{2}/h$
regime --- but also in other ways within $g\ll\nu e^{2}/h$ regime.} and assume no voltage dependence of the tunnelling amplitude $\zeta(V)$.
The use of experimentally measured $g(V)$ is then implied. This calls
for an explanation: how is it possible that the noise predictions
of the theory agree well with the coth-based formula, which in turn
agrees well with the experimental data --- yet the experimental $g(V)$
does not coincide with the theoretical $g(V)$? What is the mechanism
for agreement of a data set ``noise + tunnelling current'' with
the theory, whereas the subset ``tunnelling current'' does not agree
with the underlying theory? This question is fully applicable to Ref.~\citep{trauzettel_effect_2004}.
Reference~\citep{feldman_why_2017} avoids this criticism by deriving
its theory from statistical mechanics --- thus making no reference
to the CLL and not making any prediction for $g(V)$. Our analysis
shows that this question is resolved within the CLL framework by scaling
dimension renormalization. The renormalization provides a natural
explanation for the success of Eq.~(\ref{eq:Fano_coth}) --- given
the experimental tuning of $g(V)$ to flattness.

Further, the scaling dimension renormalization provides an explanation
for why it is possible to find flat $g(V)$ systematically, when tuning
the QPC transparency. As discussed in Sec.~\ref{subsec:discussion_sc_dim_renormalization},
interaction of the edges in the vicinity of the QPC are likely to
be the mechanism for the renormalization. Adjusting the voltage of
the gates controlling the QPC changes the positions of the edges,
changing the interaction strength, thence the scaling dimension. Note,
that other non-universalities (such as $\zeta(V)$ dependence) are
unlikely to produce flat $g(V)$.

We stress that our method proposed in this work enables the analysis
of the data at non-flat $g(V)$, significantly extending the pool
of interpretable experimental data. One can now systematically analyze
within the CLL framework the data that have typically been unexplained
and discarded.

The above discussion shows that the previously reported values of
fractional charges obtained in Fano factor experiments may often be
trustworthy. However, since the data analysis methodology may differ
from work to work, one should look into the analysis details in order
to confirm this. In the Supplemental Material \citep{SupplMat}, we
provide an example of non-trustworthy charge extraction: we perform
a naive analysis of our above experimental data using Eq.~(\ref{eq:digamma})
in the full voltage range, ignoring the CLL breakdown --- this leads
to data fits of excellent quality with wrong and temperature-dependent
values of $q$. Therefore, being pedantic about the data analysis
methodology is crucial for understanding FQH QPC tunnelling experiments.

\subsection{Violation of CLL scaling behaviour: ubiquity and possible mechanisms}

It is reasonable to ask whether the CLL breakdown observed in our
QPC is unique to the specific sample or to the filling factor of $\nu=1/3$,
or occurs ubiquitously. The data peresent in the literature do not
allow one to fully answer this question due to the lack of data for
the Fano factor at multiple temperatures in the same sample. However,
we have performed the analysis of the single-temperature data available
from recent Refs.~\citep{Kapfer2019,bartolomei_fractional_2020}
and we find signatures similar to the ones observed in our data, suggesting
that the CLL breakdown is a ubiquitous phenomenon. We present details
of this analysis in the Supplemental Material \citep{SupplMat}.

In fact, expecting the CLL picture to break down is only natural from
the theory point of view. As stated in the introduction, it only takes
an additional energy scale to destroy the CLL-predicted scaling property.
One such energy scale is the bulk gap of the FQH sample --- as soon
as one has access to such energies, one cannot focus on the edge and
should include bulk dynamics into consideration. However, we observe
the breakdown at an energy scale about 4--40 times smaller than the
bulk gap.

One can imagine a number of possible mechanisms that could lead to
the breakdown of CLL predictions with edge physics only. To mention
but a few: (i) non-linear dispersion of edge modes \citep{Nardin2023},
(ii) energy dependence of the bare tunnelling amplitude, (iii) energy-dependent
renormalization of the scaling dimension. A systematic investigation
of these mechanisms goes way beyond the scope of the present work.
In Sec.~\ref{subsec:discussion_sc_dim_renormalization} above, we
showed that mechanism (iii) predicts the correct CLL breakdown scale
--- yet we did not analyze the Fano factor beyond the breakdown scale.
In the Supplemental Material \citep{SupplMat}, we provide a toy model
mimicking mechanism (iii) and show that it can lead to the qualitative
behaviour observed: the Fano factor value at large $V$ being significantly
smaller than the quasiparticle charge $q$.

\section{Conclusions}

In this paper, we have introduced a new framework to analyze decades-old
experiments concerning quasiparticle tunnelling in the FQH effect.
Despite apparent success in determining the fractional quasiparticle
charge, the correspondence between theory and experiments across the
whole set of FQH investigations performed was unsatisfactory. Our
new methodology bridges this gap and opens the way to reconciliating
multiple quasiparticle tunnelling experiments systematically.

Our methodology is based on investigating the scaling behaviour of
the Fano factor as a function of bias voltage and system temperature.
This enables us to obtain clear signatures of CLL breakdown is some
experimental regimes and extract the quasiparticle scaling dimension
in others.

In the outlook, our methodology opens two new directions: One can
perform quantitative studies of scaling dimension renormalization
and thus get insight into the physics of mesoscopic devices. Simultaneously,
one can investigate the breakdown of CLL at quantum Hall edges in
order to get further insights into real-world, non-idealized physics
of topological matter.

\section*{Acknowledgements}

KS acknowledges illuminating discussions with Gwendal Fève, Christophe
Mora, Christian Glattli, Inès Safi, Frédéric Pierre, and Bernd Rosenow,
and thanks numerous other people working in the field for small and
big discussions over the last 15 years. K.S. acknowledges funding
by the Deutsche Forschungsgemeinschaft (DFG, German Research Foundation):
Projektnummer 277101999, TRR 183 (Project No. C01), Projektnummer
GO 1405/6-1, Projektnummer MI 658/10-2, by the German-Israeli Foundation
Grant No. I-1505-303.10/2019, as well as funding from the European
Union\textquoteright s Horizon 2020 research and innovation programme
under grant agreement No. 862683 (UltraFastNano). M.H. acknowledges
the support of the European Research Council under the European Union\textquoteright s
Horizon 2020 research and innovation programme (grant agreement number
833078). This paper was prepared with the help of \href{http://lyx.org/}{LyX}
editor.

~

\paragraph*{Note added.}

We have become aware of two experimental works addressing the question
of scaling dimension of $e/3$ quasiparticle in $\nu=1/3$ FQH effect.
The one performed in the group of F.~Pierre (A.~Veillon \emph{et
al.}, \citep{Veillon2024}) analyses data for the Fano factor with
the CLL theoretical prediction of Eq.~(\ref{eq:digamma}), yet without
scaling analysis; it finds $\delta\approx1/6$ in multiple independent
QPCs and at multiple temperatures. Another is performed in the group
of G.~Fève (M.~Ruelle \emph{et al.}, submitted to a journal); this
work uses a different method (proposed in Ref.~\citep{Jonckheere2023})
and finds $\delta\approx1/3$. The differences between $\delta\approx1/2$
we find and the scaling dimensions found in these works provide further
evidence to the non-universal effects of scaling dimension renormalization
discussed above. In light of the findings detailed in our present
work, performing a scaling analysis, in particular, examining whether
the data exhibit a similar CLL breakdown, would be important to confirm
the relevance of our findings to these works.

\section*{Methods}

\emph{Sample and device fabrication}. The sample was fabricated on
a uniform doping GaAs/AlGaAs heterostructure with a 2-dimensional
electron gas formed 118~nm underneath the surface. The transport
properties of the 2DEG showed the density $n=9.7\times10^{10}\text{ cm}^{-2}$
with mobility $\mu=4.0\times10^{6}\text{ cm}^{2}\mathrm{V}^{-1}\mathrm{s}^{-1}$
at temperature $4.2\text{\,K}$. The wet etching technique defined
electronic mesa around the QPC site, and e-gun evaporation of a metallic
stack of Ge/Ni/Au $400\text{\,nm}$ was used to form the Ohmic contacts.
The metallic gates that define the QPC were deposited on the $30\text{\,nm}$
$\mathrm{Hf}\mathrm{O}_{2}$ layer using e-beam lithography (JEOL,
JBX-9300FS) and metal evaporation of $20\text{\,nm}$ Ti/Au. The Hafnia
layer was etched using $\mathrm{B}\mathrm{Cl}_{3}$/Ar gas, and $300\text{\,nm}$
Ti/Au contact lead was deposited. All the processes except QPC patterning
used the optical lithography methods using laser writer (Heidelberg
Instruements, DWL 66+).

\emph{Measurement}. The device was placed on a mixing chamber plate
cold-finger part in a commercial wet dilution refrigerator (Leiden
Cryogenics, Minikelvin 126-TOF). Differential conductance measurements
were done via lock-in technique using a lock-in amplifier (NF corporation,
LI5655). The shot noise was amplified by ATF-34143 HEMT-based homemade
low-temperature voltage amplifier with battery-powered DC source (Standard
Research Systems, SIM928) and NF corporation SA-220F5 room temperature
voltage amplifier (second amplifying). The noise was then demodulated
by the lock-in amplifier bandwidth open method (Zurich instruments,
MFLI).

\section*{Author contributions}

N.S., Y.O., and K.S. conceived the experiment. N.S. and K.S. performed
the data analysis. T.A. and C.H. contributed to the device design,
fabrication, and measurements. V.U. has grown the sample with molecular-beam
epitaxy. N.S., T.A., C.H., M.H., Y.O., and K.S. participated in discussions
of the results and their interpretation.

\section*{Data availability}

The raw experimental data used in this work and the Mathematica code
that performs the analysis and produces the figures are available
at \href{https://doi.org/10.5281/zenodo.10840561}{https://doi.org/10.5281/zenodo.10840561}.

\bibliography{main,extras}

\clearpage{}

\part*{Supplementary material}

\beginsupplement

\section{\label{sec:suppl_g_scaling}Scaling analysis of tunnelling conductance}

CLL theory predicts scaling behaviour not only for the Fano factor,
but also for the tunnelling conductance. The tunnelling current through
a QPC is predicted to be 

\begin{equation}
I=\frac{\zeta}{\pi}(2\pi k_{B}T)^{4\delta-1}\sinh\left(\frac{qeV}{2k_{B}T}\right)\abs{\Gamma\left(2\delta+i\frac{qeV}{2\pi k_{B}T}\right)}^{2},\label{eq:current_CLL_suppl}
\end{equation}
where $\Gamma\left(x\right)$ is the Euler's gamma function and $\zeta$
encodes the bare tunnelling amplitude for quasiparticles at the QPC
(see Refs.~\citep{Wen1991,kane_transport_1992,kane_transmission_1992,Chamon1993,kane_nonequilibrium_1994}
for the original derivation and Refs.~\citep{Chamon1997,Roddaro2003,snizhko_scaling_2015}
for the expression in the notation close to the present one). This
relation can be rewritten in a scaling form:
\begin{equation}
I=\zeta T^{4\delta-1}f\left(\frac{eV}{k_{B}T}\right).
\end{equation}
Consequently, 
\begin{equation}
g=\frac{dI}{dV}=\zeta\frac{e}{k_{B}T}T^{4\delta-1}f'\left(\frac{eV}{k_{B}T}\right)=\zeta T^{4\delta-2}h\left(\frac{eV}{k_{B}T}\right).\label{eq:diff_cond_CLL_suppl}
\end{equation}

Ths specific form of functions $f$ and $h$ is not important for
the purpose of this section. What is important is that Eq.~(\ref{eq:diff_cond_CLL_suppl})
predicts that the measurements of differential conductance at different
temperatures should yield related results. Namely, when plotting $g/T^{4\delta-2}$
versus $x\equiv eV/(2\pi k_{B}T)$, the curves originating from different
temperatures $T$ should collapse on top of each other. Unlike the
Fano factor scaling considered in the main text, plotting requires
the knowledge of the scaling dimension $\delta$ (or fitting it from
the condition of optimal data collapse). We use the value of $\delta=0.54$
extracted from the Fano factor analysis. The result is presented in
Fig.~

The differential conductance data from the four highest temperatures
collapse on top of each other for $\abs x\lesssim1.5\text{--}2.5$,
just as the Fano factor data. At the same time, the data for $T=10\text{\,mK}$
does not agree with the data on the rest of the curves, unlike in
the Fano factor data. One possible explanation for such a discrepancy
is that the scaling of $g$ depends crucially on $\zeta$ being constant.
If $\zeta$ depends on $V$ or on $T$, the scaling will be ruined.
Whereas the scaling of the Fano factor is insensitive to the behaviour
of $\zeta$ as long as $g\ll\nu e^{2}/h$.

Interestingly, in our experiment, the measurements for the three lowest
temperatures were done using the same voltage on the QPC-defining
gates; whereas the voltage was set different for the two higher temperatures.
Therefore, there is no evident reason for $\zeta$ or $\delta$ to
be the same for the four highest temperatures, while different for
the lowest one. However, the scaling plot tells that this may be the
case. We have not investigated the matter further.

\begin{figure}
\centering{}\includegraphics[width=1\columnwidth]{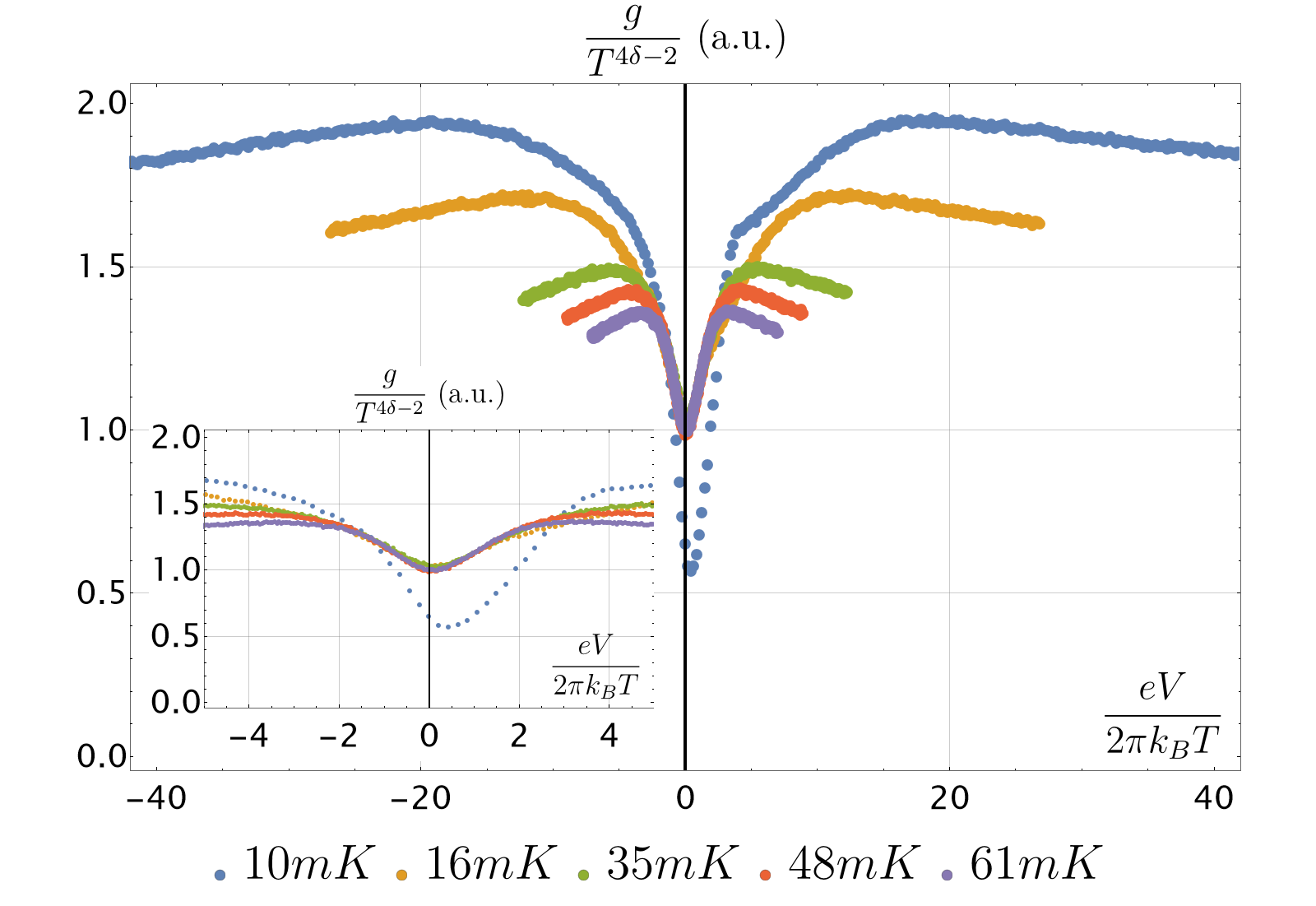}
\caption{Scaling plot of the differential conductance $g$. The data for four
highest temperatures collapse on top of each other in the same interval
of $x\equiv eV/(2\pi k_{B}T)$ as the data for the Fano factor, cf.~Fig.~\ref{fig:scaled_Fano}
of the main text. The lowest temperature data, however, does not obey
the scaling relation, which shows that the Fano factor analysis of
CLL breakdown is more reliable.}
\refstepcounter{SMfig}\label{fig:suppl_g_scaled}
\end{figure}

\section{Spurious charge and scaling dimension when ignoring CLL breakdown}

\subsection{\label{subsec:suppl_ignoring_CLL_breakdown}What happens when one
ignores CLL breakdown}

\begin{figure*}
\begin{centering}
\includegraphics[width=0.33\textwidth]{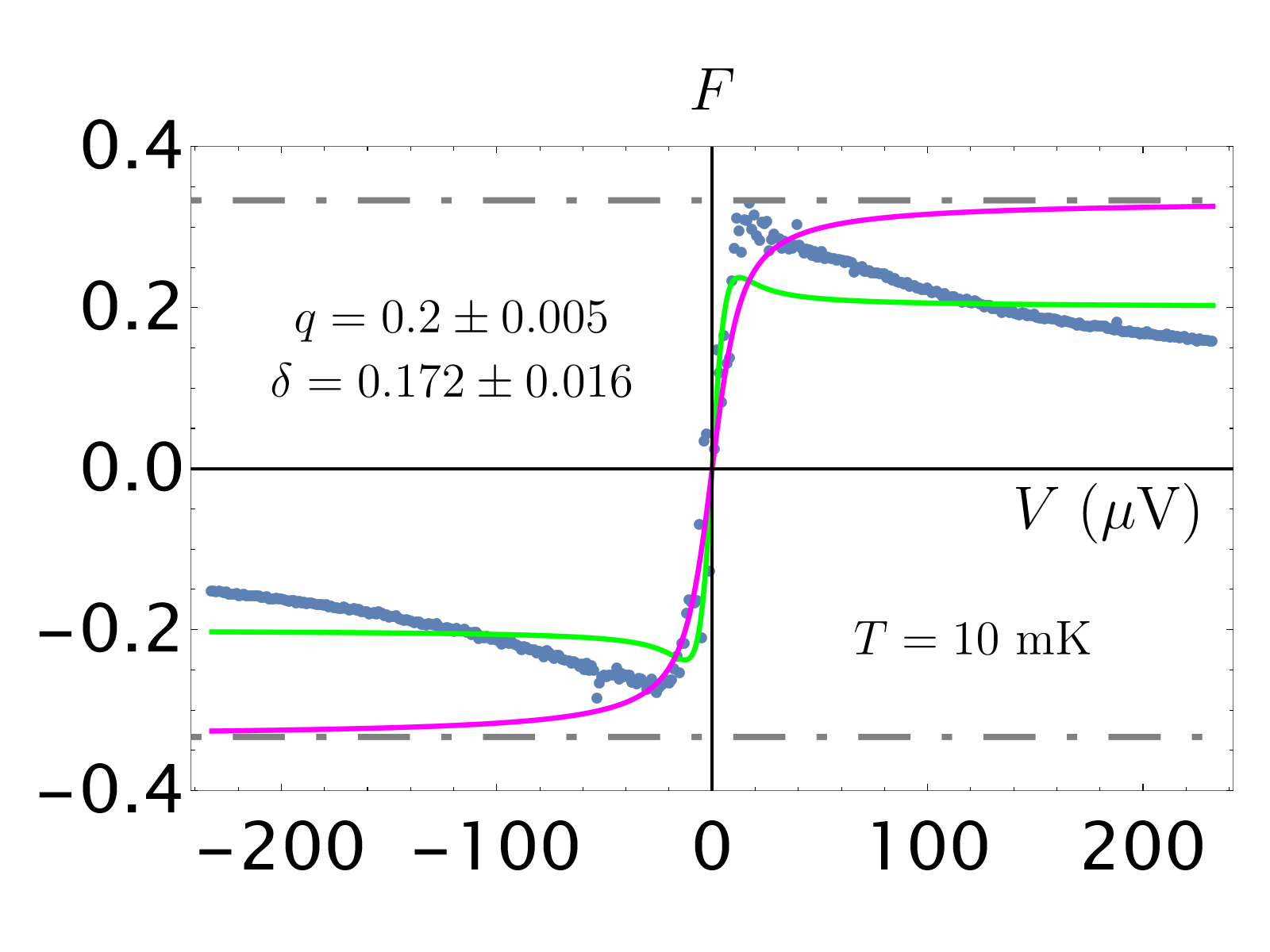}\includegraphics[width=0.33\textwidth]{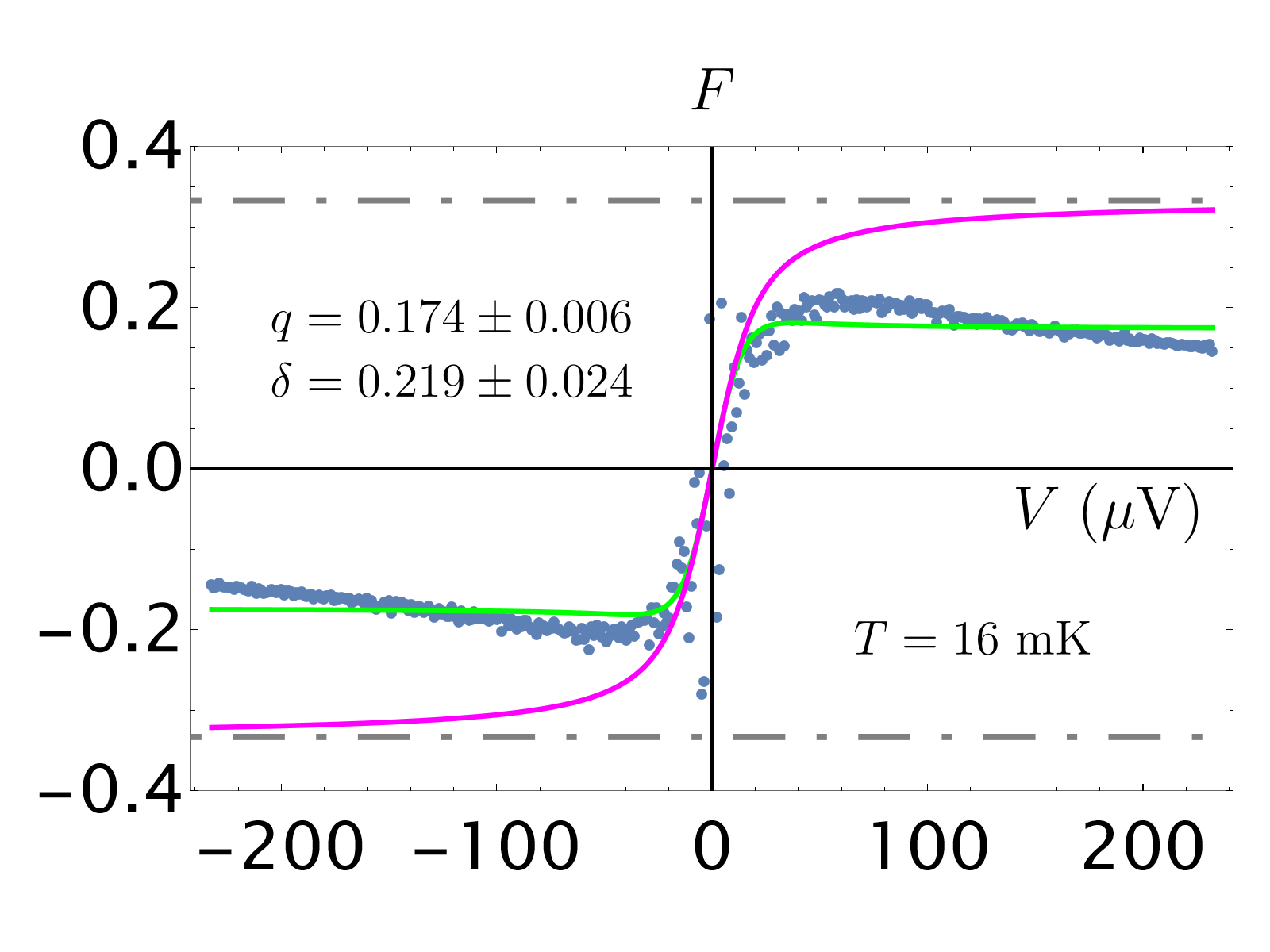}\includegraphics[width=0.33\textwidth]{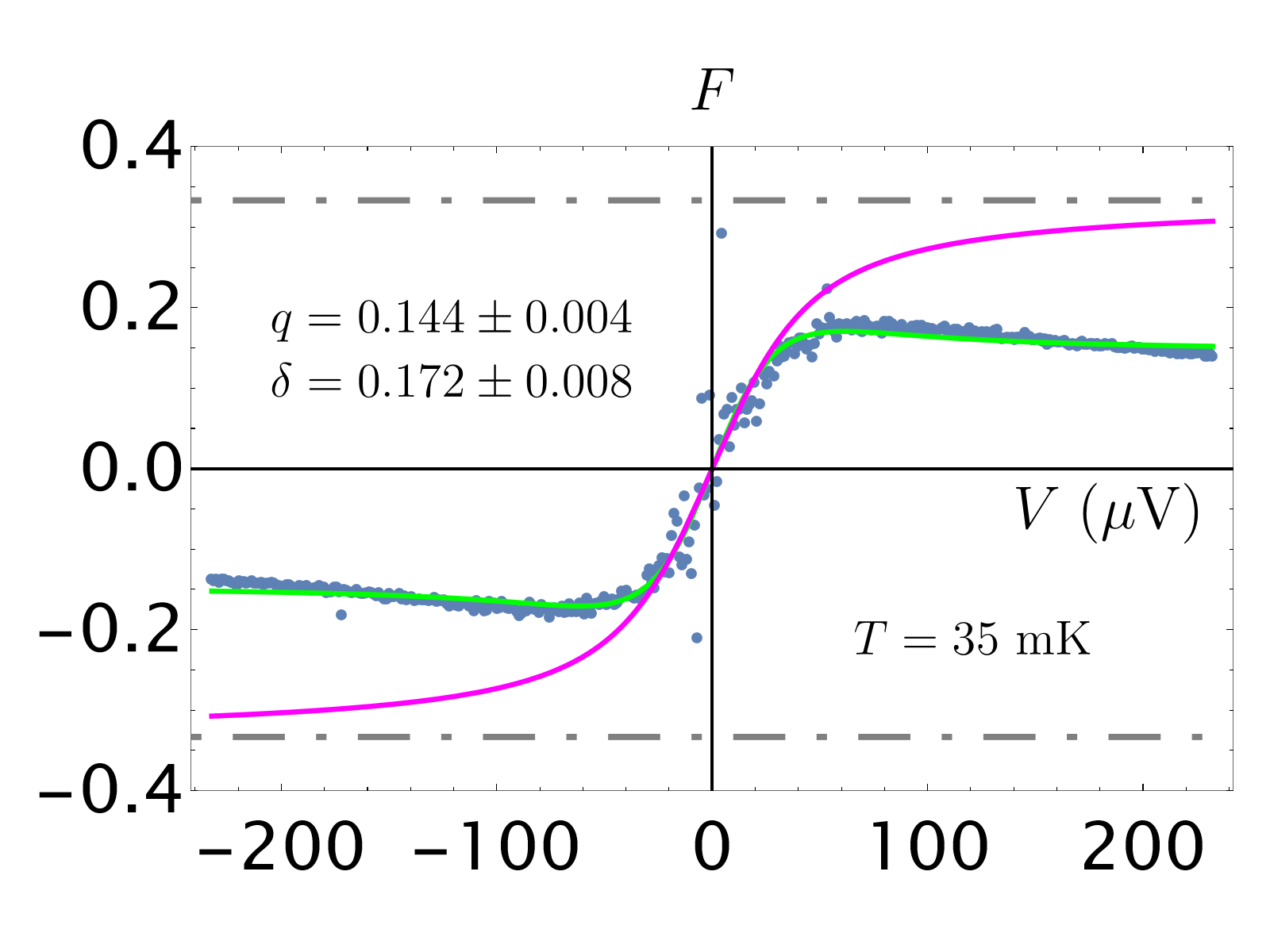}
\par\end{centering}
\begin{centering}
\includegraphics[width=0.33\textwidth]{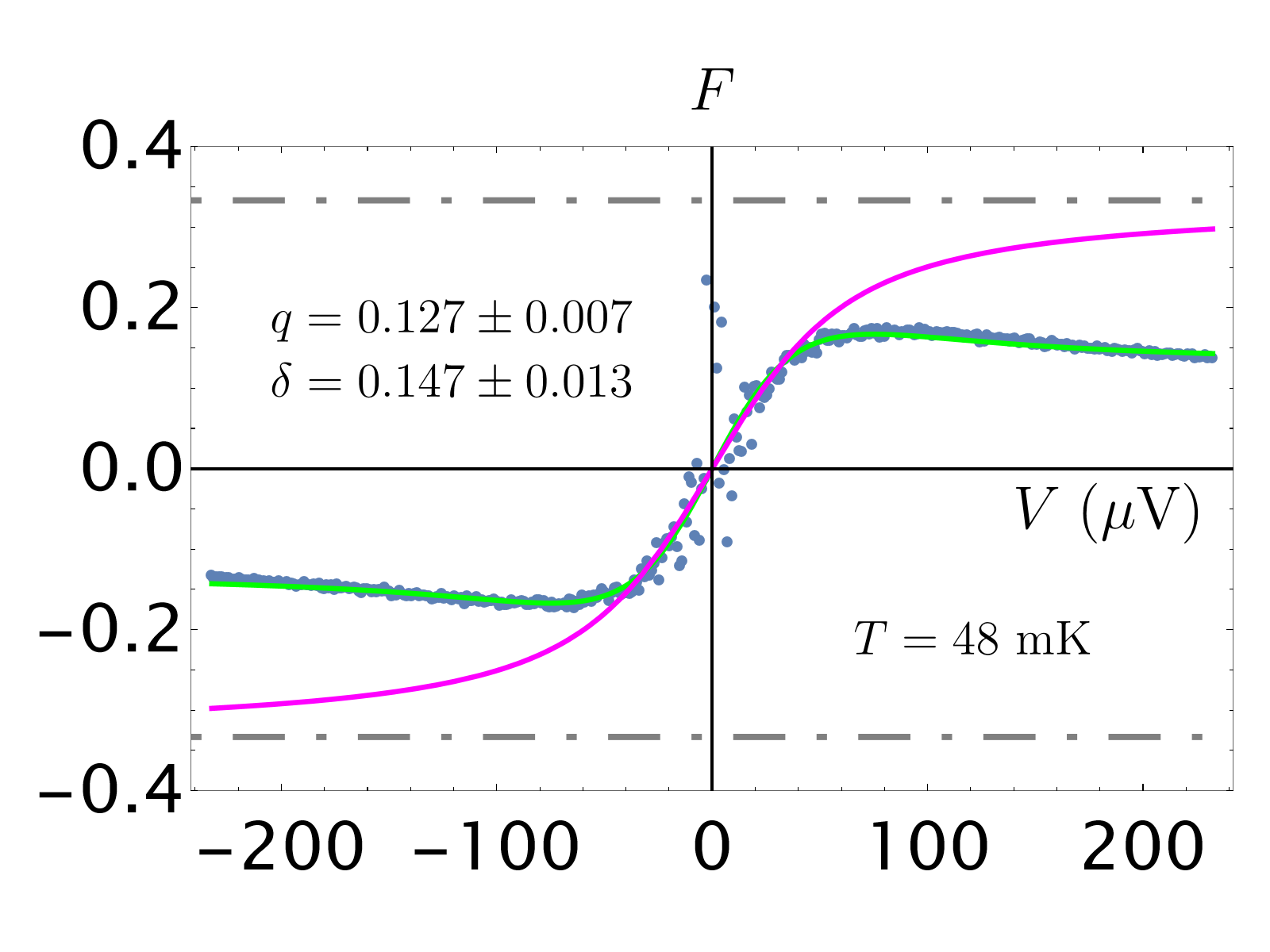}\includegraphics[width=0.33\textwidth]{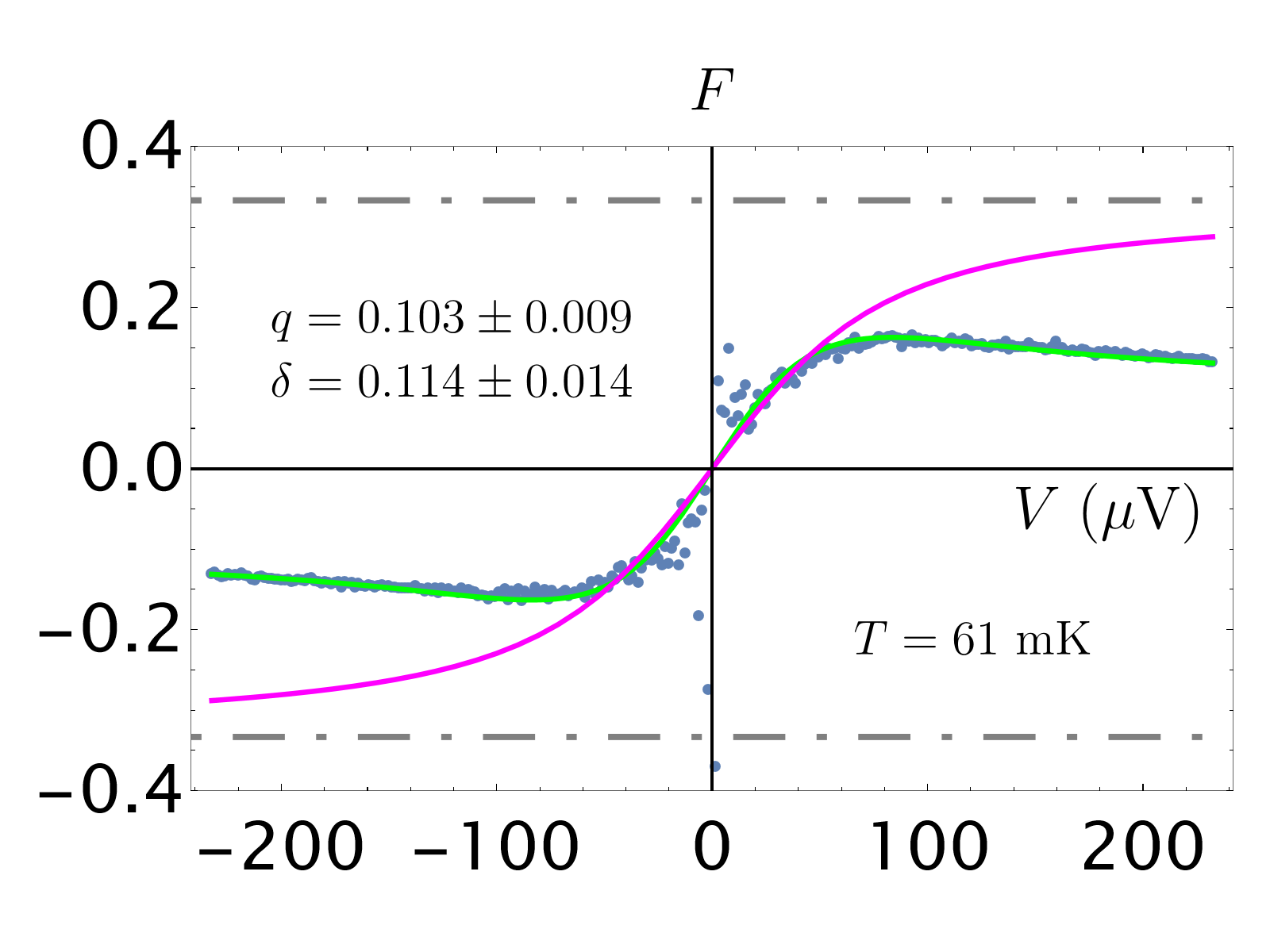}
\par\end{centering}
\centering{}\caption{Fitting our Fano factor data with the CLL prediction in the full voltage
range, ignoring the CLL breakdown. The blue points represent the data
identical to those in Fig.~(\ref{fig:raw_data}) of the main text
at the appropriate temperature $T$. The green curve is the result
of fitting the data with Eq.~(\ref{eq:digamma}) of the main text;
the extracted values for $q$ and $\delta$ with 95\%-confidence error
bars are specified in each respective plot. The magenta curve represents
Eq.~(\ref{eq:Fano_coth}) of the main text taken with $q=1/3$.}
\refstepcounter{SMfig}\label{fig:suppl_naive_fitting}
\end{figure*}

In the main text and in Sec.~\ref{sec:suppl_g_scaling}, by means
of scaling analysis comparing the data originating from different
sample temperatures, we have demonstrated that the CLL behaviour breaks
down at some energy scale. However, such analysis is not usual in
FQH tunnelling experiments and the CLL breakdown may thus go unnoticed.
Here we demonstrate that ignoring the CLL breakdown can lead to excellent
quality fits of data --- which, however, yield meaningless results.

In Fig.~\ref{fig:suppl_naive_fitting}, we fit the Fano factor data
from Fig.~(\ref{fig:raw_data}) of the main text by the CLL prediction
--- Eq.~(\ref{eq:digamma}) of the main text. We perform the fit
for each temperature separately and in the full voltage range. We
first note that the fit quality is very good, except for the lowest
temperature (yet even that is within what's conventionally tolerated
in the analysis of FQH QPC experiments). The excellent fits would
make one think that the CLL description works very well for these
data. At the same time the fitted values of $q$ are not consistent
with the value of $1/3$. Further, the fitted values of $(q,\delta)$
at different temperatures are not consistent with each other.

Such behaviour hints at the CLL breakdown we have ignored here. However,
the data at a single temperature would easily be interpreted as well-described
by the CLL with a weird value of $(q,\delta)$.

We believe that this is a likely explanation for the results of Refs.~\citep{radu_quasi-particle_2008,Lin2012,rossler_experimental_2014}
that extracted the values of $(q,\delta)$ for tunnelling current
behaviour. Indeed, those works have made ad-hoc adjustments to the
CLL formulas to account for what may be CLL breakdown --- and then
performed fitting in the full data range, all at a single temperature.

\subsection{Signatures of CLL breakdown in other works}

In the available literature, the noise data in FQH QPCs is typically
measured at a single temperature. This makes it complicated to firmly
establish the CLL breakdown. However, hints at the CLL breakdown can
be often seen. We demonstrate this here by considering the supplementary
data from Refs.~\citep{Kapfer2019,bartolomei_fractional_2020} (each
of these works focused on a different matter, however, provided the
data for the standard measurements as a supplement).

\begin{figure}
\centering{}\includegraphics[width=1\columnwidth]{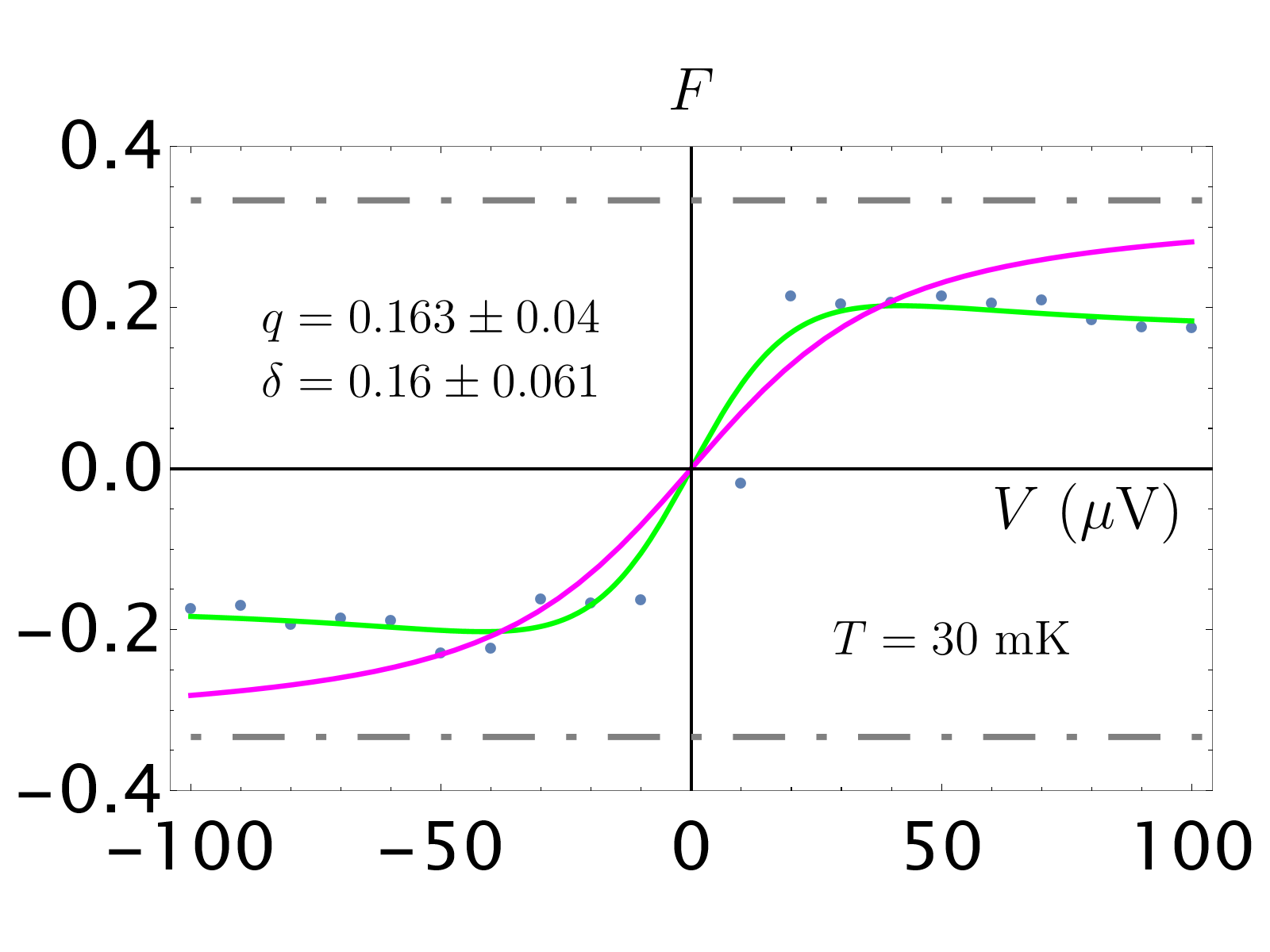}
\caption{Fitting the Fano factor data calculated based on Fig.~S3(c) of the
SM of Ref.~\citep{bartolomei_fractional_2020}; the data relate to
$\nu=1/3$ and correspond to transmission $T_{1}=0.08$ ($g=0.08\nu e^{2}/h$
in our notation) and temperature $T=30\text{ mK}$. The magenta curve
represents Eq.~(\ref{eq:Fano_coth}) of the main text taken with
$q=1/3$ and shows the size of errors tolerated in the standard analysis
of FQH QPC experiments. The fit with the CLL prediction (green) ---
Eq.~(\ref{eq:digamma}) of the main text --- describes the data
much better, yet yields unexpected values of $q$ and $\delta$. Remark
the similarity of the fit and the values to the $35\text{ mK}$ data
in Fig.~\ref{fig:suppl_naive_fitting}.}
\refstepcounter{SMfig}\label{fig:suppl_naive_fitting-feve}
\end{figure}

Figure~S3 of the supplemental material of Ref.~\citep{bartolomei_fractional_2020}
provides the data on the behaviour of two ``injection'' QPCs used
in their ``anyonic collider'' experiment. The data correspond to
the filling factor $\nu=1/3$ at temperature $T=30\text{ mK}$ and
various levels of $g/(\nu e^{2}/h)$ (8 data sets in total). In Fig.~\ref{fig:suppl_naive_fitting-feve},
we analyze one representative data set. First note that the standard
coth formula (Eq.~(\ref{eq:Fano_coth}) of the main text) agrees
with the data rather loosely --- which shows that the standard analysis
tolerates rather large errors.\footnote{Such tolerance is acceptable when one seeks to establish the existence
of fractional quasiparticles, as was originally the aim of such experiments
\citep{de-picciotto_direct_1997,saminadayar_observation_1997}: within
such tolerance, $F\approx0.2\text{--}0.3$ is clearly distinct from
the $F=1$ characterizing non-interacting electrons. However, this
level of tolerance clearly does not enable quantitative characterization
of fractionalized excitations.} The magnitude of the errors can be seen in the original Figure~S3(c)
of Ref.~\citep{bartolomei_fractional_2020} --- even though the
standard way of plotting the noise, and not the Fano factor, makes
the errors appear smaller, they are still quite noticeable.

The fit of the same data by the CLL prediction shows the behaviour
similar to that discussed in Sec.~\ref{subsec:suppl_ignoring_CLL_breakdown}
--- the fit quality is excellent, the extracted values of $(q,\delta)$
are nonsensical. In our view, this is a strong signature of the CLL
breakdown taking place. In the code we provide with this paper (\href{https://doi.org/10.5281/zenodo.10840561}{https://doi.org/10.5281/zenodo.10840561}),
we give the reader an opportunity to perform the same analysis of
other data sets from Fig.~S3 of the supplemental material of Ref.~\citep{bartolomei_fractional_2020}.

\begin{figure}
\centering{}\includegraphics[width=1\columnwidth]{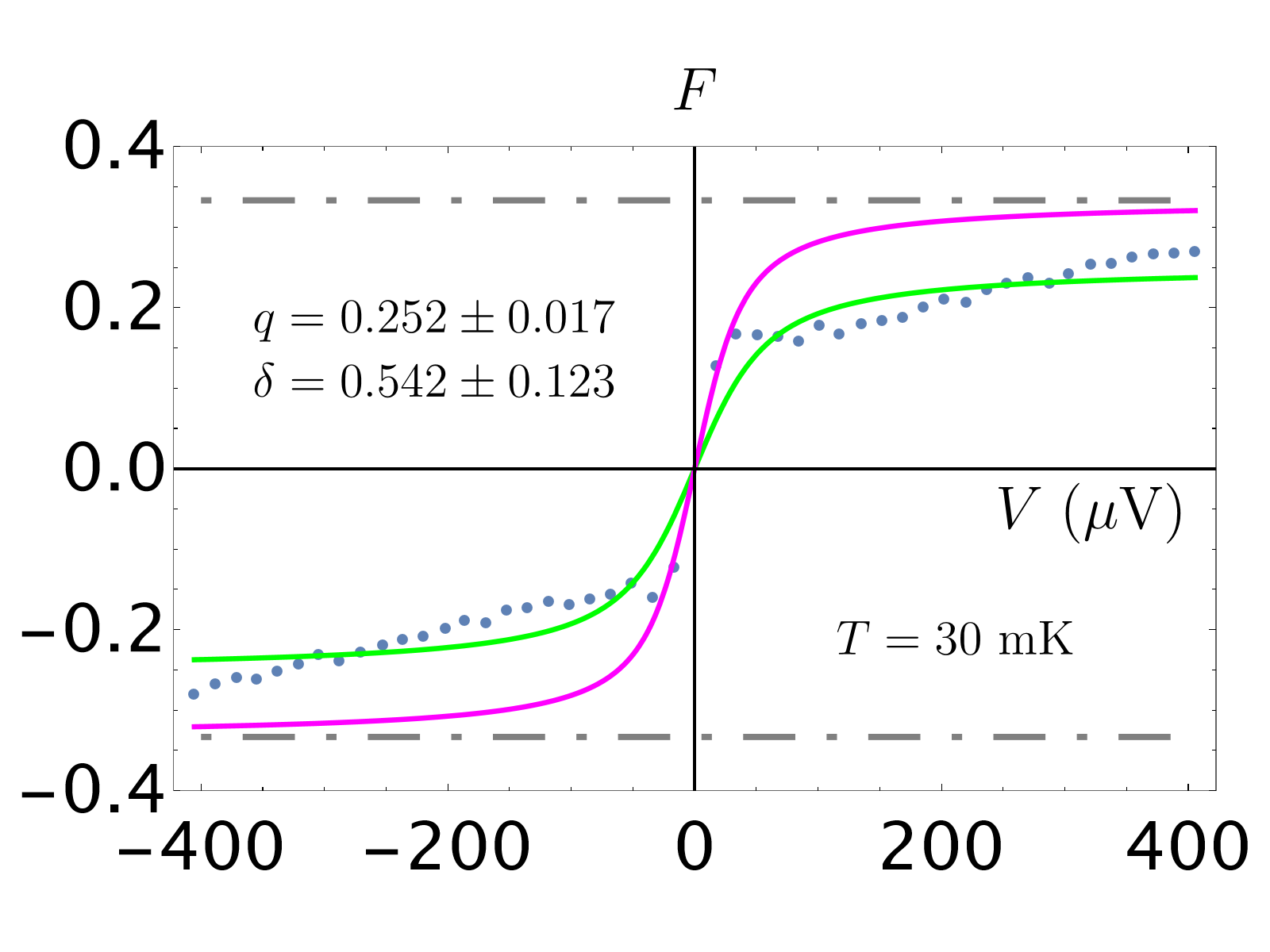}
\caption{Fitting the Fano factor data calculated based on Figs.~2(B) and S3
of Ref.~\citep{Kapfer2019}; the data relate to $\nu=1/3$ edge mode
of the $\nu=2/5$ edge; the transmission level $g/(1/3\times e^{2}/h)\lesssim0.05$
and the temperature $T=30\text{ mK}$. The magenta curve represents
Eq.~(\ref{eq:Fano_coth}) of the main text taken with $q=1/3$ and
shows the size of errors tolerated in the standard analysis of FQH
QPC experiments. The fit with the CLL prediction (green) --- Eq.~(\ref{eq:digamma})
of the main text --- describes the data much better, yet yields unexpected
values of $q$ and $\delta$. Note that the Fano factor continues
growing at the voltages where saturation is expected.}
\refstepcounter{SMfig}\label{fig:suppl_naive_fitting-glattli}
\end{figure}

We next analyse the zero-frequency noise data from Ref.~\citep{Kapfer2019}.
In this work, the $1/3$ edge mode was part of the $\nu=2/5$ edge.
Also, the voltage range of the data is much larger than typically
used. We analyze the data in Fig.~\ref{fig:suppl_naive_fitting-glattli}.
Again, the fit by the CLL prediction is much better that the description
by the coth-based phenomenological formula\footnote{Interestingly, the deviation from the coth-based prediction for noise
is hardly noticeable in Fig.~2(B) of Ref.~\citep{Kapfer2019}. The
origin of this discrepancy with our plot is unclear to us. It may
be another instance when plotting the noise masks the errors that
are evident when the Fano factor is considered.} and the extracted values of $(q,\delta)$ are hardly consistent with
the ones expected. Yet, here the fit quality is much lower than what
we have seen in Figs.~\ref{fig:suppl_naive_fitting} and \ref{fig:suppl_naive_fitting-feve}.
Noticeably, the Fano factor keeps growing when the saturation is expected
both from the CLL fit and the coth formula. The behaviour difference
to our data in Fig.~\ref{fig:suppl_naive_fitting} and the data of
Ref.~\citep{bartolomei_fractional_2020} in Fig.~\ref{fig:suppl_naive_fitting-feve}
may stem from the fact that here $1/3$ mode of the $2/5$ edge is
investigated. Nevertheless, we interpret the results shown in Fig.~\ref{fig:suppl_naive_fitting-glattli}
as a hint that the CLL breakdown happens in this system too.

Given that above analysis and the results of Refs.~\citep{radu_quasi-particle_2008,Lin2012,rossler_experimental_2014}
(which measured tunnelling current only), it appears that CLL breakdown
is a ubiquitous behaviour, which has often gone unnoticed, leading
to confusing results in the literature.

\section{Energy-dependent scaling dimension renormalization --- toy model}

Our results in the main text indicate a renormalization of the scaling
dimension towards $\delta\approx1/2$ at low energies, with some breakdown
of CLL theory at higher energies. Among other features, this results
in a deviation from the predicted asymptotic tendency of the Fano
factor towards quasiparticle charge at high voltages.

Inspired by these results and by Landauer-Buttiker-Imry scattering
theory \citep{blanter_shot_2000}, we present here a toy model of
an energy-dependent scaling dimension. We show that a model of this
form can recreate qualitative features that we have obtained experimentally.
We make no explicit assumption as to the source of this crossover
energy scale; however, it has been suggested that the geometry of
the QPC, and in particular its width, may affect the tunnelling current
\citep{yang_influence_2013,zucker_edge_2015}. An additional potential
energy scale which could be introduced into the system and distinguish
high- and low-energy behaviour is the energy associated with Coulomb
interactions between the edges.

We begin with the general expressions for the tunnelling current and
noise using Landauer-Buttiker-Imry scattering theory. We consider
a system of two reservoirs, denoted here by $R$ and $L$, with a
reflection coefficient of $R(E)$ between them. To the leading order
in $R(E)$, the tunnelling current $I$ and the excess\footnote{As in the main text, we define the excess noise as the noise added
by a non-zero voltage, rather than as the noise added by the partitioning.} tunnelling current noise $S'$ are given by \citep{blanter_shot_2000} 

\begin{align}
I= & \frac{qe}{h}\int dE|R(E)|^{2}\left(f_{R}(E)-f_{L}(E)\right)\label{eq:LandauerButtiker_current}\\
S'= & \frac{2\left(qe\right)^{2}}{h}\int dE|R(E)|^{2}\left(f_{R}(E)-f_{L}(E)\right)^{2},\label{eq:LandauerButtiker_noise}
\end{align}
 where $f_{\alpha}(E)=\left[\exp{\left((E-qeV_{\alpha})/(k_{B}T)\right)}+1\right]^{-1}$
is the Fermi-Dirac distribution. Note that we have heuristically replaced
the electron charge with the quasiparticle charge, $qe$. Also note
that $S'$ corresponds to the noise in a two-terminal QPC experiment,
not the 4-terminal one as considered in the rest of the work. This,
however, suffices for our purpose in this section, as the expected
value of the Fano factor in the 2-terminal experiment is the same.
Indeed, when $T=0$ or for $R(E)\approx\mathrm{const}$ and $e\abs{V_{L}-V_{R}}\gg k_{B}T$,
the dominant contribution to the above integrals comes from the regions
where $f_{R}(E)-f_{L}(E)=\sgn\,(V_{R}-V_{L})=\pm1$ --- so, the integrals
coincide up to a sign, producing $F'=S/(2eI)=q$. In what follows,
we take $V_{L}=0$ and $V_{R}=V$.

Further, energy-dependent $|R(E)|^{2}\propto\left(\abs E+E_{0}\right)^{4\delta-2}$
can be a toy model for mimicking the CLL behaviour; $E_{0}$ is a
low-energy cutoff (which can be discarded in Eqs.~(\ref{eq:LandauerButtiker_current}--\ref{eq:LandauerButtiker_noise})
if $\delta>1/4$). At $T=0$ for $V\gg E_{0}$, such $|R(E)|^{2}$
produces $dI/dV\propto V^{4\delta-2}$ in agreement with the CLL prediction
and $S'=2qeI$, so that $F'=q$.

\begin{figure}
\centering{}\includegraphics[width=1\columnwidth]{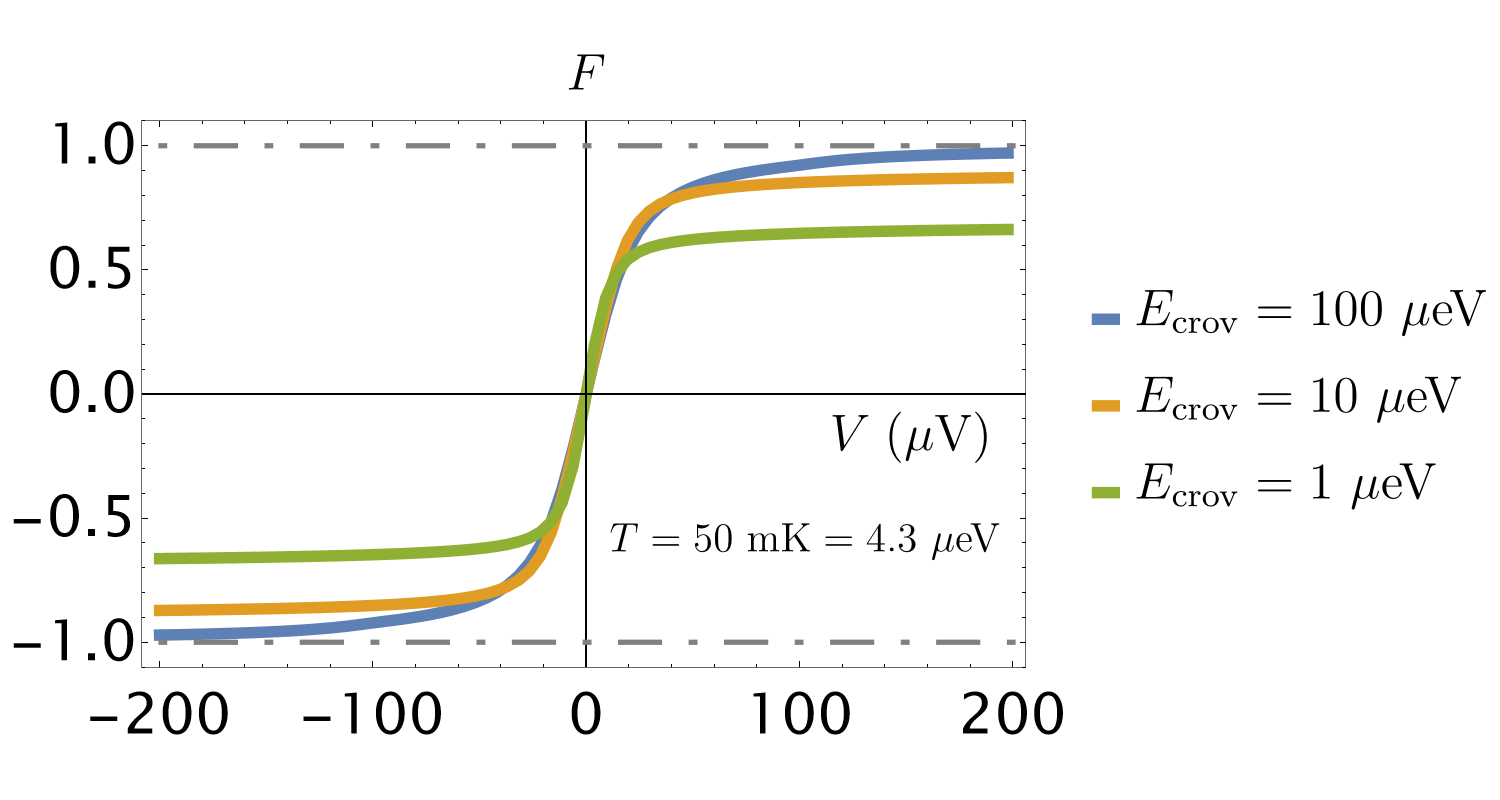}
\caption{Fano factor for a toy model (\ref{eq:LandauerButtiker_current}--\ref{eq:bifurcation})
mimicking energy-dependent scaling dimension renormalization. The
large-$V$ value of the Fano factor clearly can differ from $q=1$,
depending on the energy scale $E_{\mathrm{crov}}$, where the scaling
dimension changes.}
\label{fig:toy_model}
\end{figure}

In order to model energy-dependent renormalization of the scaling
dimension, we take the transmission coefficient to be 
\begin{equation}
|R(E)|^{2}=\begin{cases}
R_{0}\left(\frac{\abs E+E_{0}}{E_{\mathrm{crov}}}\right)^{4\delta_{1}-2}, & E\leq E_{\mathrm{crov}}\\
R_{0}\left(\frac{\abs E+E_{0}}{E_{\mathrm{crov}}}\right)^{4\delta_{2}-2}, & E>E_{\mathrm{crov}}
\end{cases}.\label{eq:bifurcation}
\end{equation}
Here $E_{\mathrm{crov}}$ is a crossover energy scale, which separates
the low-energy behaviour determined by scaling dimension $\delta_{1}$
and the high-energy behaviour determined by scaling dimension $\delta_{2}$.

In Fig.~\ref{fig:toy_model}, we display how the size of $E_{\mathrm{crov}}$
affects the Fano factor. We choose $\delta_{1}=0.54$, inspired by
the results of the main-text analysis, and $\delta_{2}=1/6$, inspired
by the theoretically predicted value for the $\nu=1/3$ Laughlin state
in the absence of renormalization. The value of the charge is chosen
to be $q=1$ for convenience. We plot the Fano factor as a function
of voltage, with a temperature of $T=50\text{ mK}$.

As can be seen in the figure, the value of $E_{\mathrm{crov}}$ greatly
affects the asymptotic tendencies of the Fano factor. For $E_{\mathrm{crov}}=100\text{ \ensuremath{\mu}eV}\gg k_{B}T$,
the Fano factor asymptotically tends towards $F=q\,\textrm{sgn}V$,
as predicted by the CLL theory. For smaller $E_{\mathrm{crov}}$,
however, the asymptotic value of the Fano factor is significantly
reduced. 

We emphasize that this is a toy model, and not a proper CLL calculation.
However, this toy model demonstrates that energy-dependent renormalization
of the scaling dimension may explain the Fano factor's unexpected
behaviour at large $V$.
\end{document}